\documentclass[aps,prf,superscriptaddress,longbibliography]{revtex4-2}
\usepackage{amsmath,amssymb,graphicx,float,epstopdf}
\usepackage{dcolumn}
\usepackage{bm}
\usepackage{color}
\usepackage{setspace}

\usepackage{tikz}
\usetikzlibrary{arrows.meta,bending,positioning,intersections}
\usepackage{pgfplots}
\usepackage[all]{xy}

\begin{document}

\title{Dynamics of pivoted slider bearings} 

\author{Andrew Wilkinson}
\affiliation{School of Mathematics and Statistics, The Open University, Milton Keynes MK7 6AA, United Kingdom}
\author{Marc Pradas}
\email[]{marc.pradas@open.ac.uk}
\affiliation{School of Mathematics and Statistics, The Open University, Milton Keynes MK7 6AA, United Kingdom}
\author{Michael Wilkinson}
\email[]{m.wilkinson@open.ac.uk}
\affiliation{School of Mathematics and Statistics, The Open University, Milton Keynes MK7 6AA, United Kingdom}

\date{\today}

\begin{abstract}
We obtain the full equations of motion for a 
wide, pivoted, slider bearing. These are used to 
review the choice of the optimal position for the pivot point, to 
discuss its response to time-dependent sliding velocity, and to 
determine the stability of the motion. 
The case of an abrupt acceleration of the slider, which results in a large 
transient increase of the resistive force, is surprisingly complicated.
We also discuss a general  \emph{transversion} formula for changing 
the dependent variables in Stokes flow problems.
\end{abstract}

\maketitle 

\section{Introduction}
Slider bearings are used to reduce friction between two solid objects. 
A thin film of lubricant viscous fluid between the surfaces allows them to slide over each 
other with relatively little friction, while the pressure of the fluid allows them to remain 
separated despite a large load perpendicular to the surfaces \cite{Bat67,Sze98, Or+97}. 
Slider bearings occur in various forms: here we consider a simple slider which facilitates linear motion 
of one flat surface over a fixed, flat base. Similar principles are applicable to thrust bearings,
which support a force directed along a rotating axis, and journal bearings, where the 
force is perpendicular. Slider bearings are important element of mechanical engineering, 
finding applications in almost all systems which involve large moving parts. 

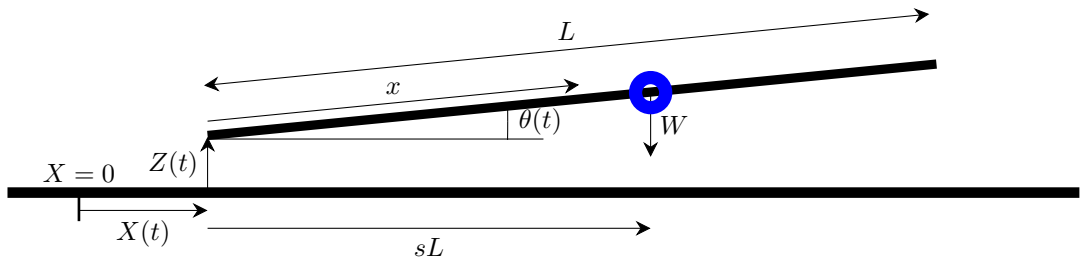
\begin{figure*}
\centering
\begin{tikzpicture}[scale=0.95]
\path[draw,line width=4pt](2,0)--(17,0);
\path[draw,line width=3.5pt](4.8,0.8)--(15,1.8);
\draw[-{Stealth[length=2mm, width = 2mm]}] (4.8,0) -- (4.8,0.78) node[pos=.5,left] {$Z(t)$};
\draw[-] (4.8,0.75) -- (9.5,0.75) node[pos=0.65,above,yshift=-0.06cm] {$\qquad \qquad \qquad \qquad \ \ \theta(t)$};
\draw[-{Stealth[length=2mm, width = 2mm]}] (4.8,1) -- (10,1.51) node[pos=.5,above] {$x$};
\draw[-{Stealth[length=2mm, width = 2mm]}] (3,-0.25) -- (4.8,-0.25) node[pos=.5,below] {$X(t)$};
\draw[-{Stealth[length=2mm, width = 2mm]}] (11,1.4) -- (11,0.5) node[pos=.5,right] {$W$};
\draw[-{Stealth[length=2mm, width = 2mm]}] (4.8,-0.5) -- (11,-0.5) node[pos=.5,below] {$sL$};
\draw[{Stealth[length=2mm, width = 2mm]}-{Stealth[length=2mm, width = 2mm]}] (4.75,1.5) 
-- (14.9,2.5) node[pos=.5,above] {$L$};
\draw[line width=1pt] (3,-0.4)--(3,0) node [pos=1.0,above]{$X=0$};
\draw[] (9,1) arc (0:1:15);
\draw[] (9,1) arc (0:-1:15);
\draw[blue,line width=5pt] (11,1.4) circle (6pt);
\end{tikzpicture}
\caption{
The slider bearing has length $L$ and its pivot point (the blue circle) is displaced from the 
left-hand (trailing) edge by $sL$. The bearing is assumed to be sufficiently deep 
that motion of the fluid in the perpendicular direction to the plane of this diagram 
can be neglected. The pivot supports a weight per unit depth $W$. Motion of 
the upper element is to the right, with specified speed $v(t)$, is opposed 
by a horizontal viscous drag force per unit depth equal to $F_x(t)$.  
At a time $t$, the angle of the plate is $\theta(t)$, the gap at the left-hand edge is $Z(t)$ 
and the horizontal displacement of the left-hand edge is $X(t)$. 
(Sketch is not to scale).
}
\label{fig: 1}
\end{figure*}

The operation of slider bearings depends upon maintaining the film of lubricant. 
This can be effected by the sliding surface being at an angle to the fixed plate so that 
the motion entrains lubricant into the gap. One very effective method to ensure that the 
plate is at a suitable angle is by using a  pivoted slider bearing~\cite{Mic50}. 
This allows the angle $\theta$ between the two 
surfaces to vary in response to varying the sliding speed $v$ of the surface. 
In this configuration, the slider bearing has a pivot point that carries all the vertical load, as 
illustrated in figure \ref{fig: 1}. This system was discussed in the book by 
Michell~\cite{Mic50}, but the techniques that were used there were limited to 
considering steady-state solutions with a constant sliding velocity $v$. 
However, in general both the angle $\theta$ and the gap at the trailing edge 
of the slider bearing, $Z$, are time-dependent, and we should consider the 
case where $v$ has a specified time-dependence. In this paper we obtain 
and analyse the equations of motion for $\theta$ and $Z$ when the slider 
velocity $v$ is time-dependent. 

In the following,  we assume that the length of the slider surface is $L$, 
and that the position of the pivot point  is at a distance $sL$ in front of the trailing edge 
of the bearing surface (see figure \ref{fig: 1}). In steady motion, the optimal position $s$ of the pivot 
might be taken to be that which minimises the resistive force $F_x$, with other variables fixed. 
This question was addressed in Michell's book~\cite{Mic50}. His results show that the optimal 
pivot position, $s^\ast$, is independent of $v$. Also, the steady-state values of $\theta$, 
$Z$ and $F_x$ are all proportional to $\sqrt{v}$, as will be explained 
in section \ref{sec: 3}. 

In our formulation, we introduce convenient dimensionless variables 
\begin{equation}
\label{eq: 1.1}
\eta=\frac{\theta L}{Z}
\ ,\ \ \ 
\xi=\frac{Z}{L}
\ ,\ \ \ 
\lambda=\ln \xi\,
\end{equation}
which we shall term, respectively,  the aspect parameter, the gap parameter, and the 
logarithmic gap parameter. Because the steady-state values of both $\theta$ and $Z$ are proportional to 
$\sqrt{v}$, in the steady state the aspect parameter $\eta$ approaches an 
equilibrium value $\eta_0$ which is independent of $v$ (and which is, 
in fact, a function only of the pivot point parameter, $s$). Our results on the steady 
state are shown to be compatible with those in Michell's book. In particular, we
consider the optimal configuration for a slider bearing operated in the steady state, 
by minimising the resistance force $F_x$ at a given value of the speed $v$. 
Our analysis shows that this minimum as achieved at a value of the pivot 
parameter $s^\ast\approx 0.3904$, with aspect parameter $\eta_0\approx 2.0713$.
These values are consistent with values quoted in \cite{Mic50}: see table IV, p.~81 of Michell's book.

We use our equations of motion to extend the analysis of pivoted slider bearings 
beyond steady conditions. We make a linear stability analysis, and find that the 
steady-state solutions investigated by Michell are always stable. This implies
that, if the bearing is operating close to its steady-state condition and we make a 
small change to $v$, the bearing will converge to its new equilibrium. 
The timescale for relaxation is of order $L/v$, as might be anticipated 
on physical grounds. The new equilibrium has the same value of $\eta_0$, 
but the value of $\eta$ is not constant during the transition.

If the slider bearing is abruptly stopped, we show that there is a solution 
for which the gap parameter decreases as $\xi \sim t^{-1/2}$, with a different equilibrium 
value of  $\eta$, denoted by $\eta_-$. The value of $\eta_-$ depends only upon the 
pivot position, $s$. We find that these sinking solutions are not stable when the pivot 
point is too close to the end of the bearing. In cases where this solution is unstable, 
we hypothesise that one end of the slider contacts the surface in finite time.

The most technologically significant case of non-steady motion is 
when the bearing is subject to an abrupt acceleration, because the resistive force 
will be greatly increased during the transient when the gap parameter $\xi(t)$ is 
increasing. We analyse the case where the velocity is increased to a constant value
$v$, and investigate how $\eta$ and $\xi$ evolve towards their new equilibrium values. 
We find that the dynamics of  
$\eta(t)$ and $\xi(t)$ during this transient are complex. If the equilibrium 
gap parameter for the increased velocity is much greater than the initial value, 
we show that the value of $\eta$ becomes very small before reaching its equilibrium 
value, and that the duration of the transient is much greater than $L/v$. 
We characterise the effect of the acceleration 
by estimating the excess energy dissipated, $\Delta E$, 
during acceleration to speed $v$ from an initial condition characterised by 
coordinates $\eta_{\rm in}$, $\xi_{\rm in}$. If $F_x(t)$ is the resistive force 
at time $t$ after initiating the speed increase, and $F_x(\infty)$ denotes 
the equilibrium value of this force, then the excess energy dissipated is
\begin{equation}
\label{eq: 1.2}
\Delta E=\int_0^\infty {\rm d}t\ \left[F_x(t)-F_x(\infty)\right]v(t)
\ ,
\end{equation}
which we evaluate as a function of the velocity $v$, the initial coordinates 
$\xi_{\rm in}$, $\eta_{\rm in}$, and the pivot position parameter $s$.

In an earlier paper~\cite{Wil+23}, we used the Reynolds equations of 
lubrication theory \cite{Rey86} to determine the equations of motion 
for a flat plate settling onto a flat, horizontal surface. The system has three 
degrees of freedom, namely the gap $Z(t)$, angle $\theta(t)$ and horizontal 
displacement $X(t)$. The equations of motion connect a three-dimensional, 
dimensionless, generalised velocity $\dot {\mbox{\boldmath$x$}}$ to a 
dimensionless generalised force vector $\mbox{\boldmath$f$}$, via a resistance 
matrix ${\bf B}$, i.e., we have $\mbox{\boldmath$f$}={\bf B}\dot{\mbox{\boldmath$x$}}$.
For the equations of motion for the slider bearing considered in this paper, 
the same dynamical elements appear, 
but the independent variables are a mixture of forces and velocities. In section \ref{sec: 2}
we discuss a general expression for what we shall term the 
\emph{transversion} of elements between the 
independent and dependent components in a linear equation of motion. We use this 
formula to obtain the equations of motion for the slider bearing. 

The rest of the paper is organised as follows.
In section \ref{sec: 3} we discuss the steady-state solution considered
by Michell, which can be obtained from our general form of the equations of motion.  
In section \ref{sec: 4} we discuss the local stability properties of the steady-state 
solution, showing that these solutions are all stable. We determine the relaxation 
rates, showing that they are of order $L/v$. In section \ref{sec: 5} we consider the settling of the 
bearing when $v=0$, showing that there are solutions with $\eta$ 
approaching a constant, $\eta_-$, and with $\xi\sim t^{-1/2}$. 
We show that this solution is only stable when the pivot position 
is  in an interval close to the centre of the plate. 
Section \ref{sec: 6} discusses the case where $v$ is suddenly 
increased by a large factor,  and estimates the excess energy dissipated.  
Section \ref{sec: 7} is a brief conclusion and summary.

\section{Equations of motion}
\label{sec: 2}

The problem analysed in this paper is sketched in Fig.~\ref{fig: 1}. 
A slider bearing of length $L$ has a pivot point located at a distance $sL$ 
from the left-hand edge of the bearing. We consider the 
case where the slider bearing is moving with a specified speed $v(t)$.  
We assume that the slider bearing 
is immersed in a viscous fluid of viscosity $\mu$ and 
constant  density $\rho$,  and we model the motion of 
the vertical gap $Z(t)$ between the left-hand edge of the bearing and the 
lower solid surface, and the angle $\theta(t)$ between the slider bearing and the horizontal. 

The problem sketched in Fig.~\ref{fig: 1} is described in terms of the same 
dynamical variables as the problem of  a flat plate settling onto a flat 
horizontal surface  under the conditions of lubrication theory. This problem 
was analysed in \cite{Wil+23}, where the equations of motion were expressed 
in dimensionless coordinates, $\eta=\theta L/Z$, $\xi =Z/L$, $\lambda=\ln \xi$, 
and a further dimensionless variable $\zeta=X/L$, where $X$ is the horizontal displacement 
of the plate.  The equations of motion in \cite{Wil+23} also introduced a dimensionless  time variable. 
In this work we choose the dimensionless time $\tilde t$, defined by
\begin{equation}
\label{eq: 2.1}
\tilde t=\frac{W}{\mu L} t
\end{equation}
where $W$ is the downward force per unit depth on the bearing, see Fig.~\ref{fig: 1}. 
(In the following, reference to a  force  is understood to be force per unit depth.) 
The motion of this system can be described by a generalised 
velocity vector, with components $(v,{\rm d}\lambda/{\rm d}t,{\rm d}\eta/{\rm d}t)$, 
which can be related to a generalised force vector, $(F_x,F_z,G)$, where 
$F_x$ is the horizontal force, $F_z$ the vertical force, and $G$ the moment of the vertical 
force about the trailing edge of the bearing. If the bearing supports a weight $W$, 
then $F_z=W$ and $G=WsL$. It will be convenient to express these vectors in dimensionless
form, by using the dimensionless time $\tilde t$ given by Eq.~\eqref{eq: 2.1} 
to define a dimensionless velocity vector, 
and defining a dimensionless force vector by dividing the elements $F_z$ by $W/\xi^2$.   
Hence, we have the dimensionless generalised force vector $\boldsymbol{f}$ 
and velocity vector $\dot{\boldsymbol{x}}$ given by
\begin{equation}
\label{eq: 2.2}
\left(\begin{array}{c}
 f_1  \\[6pt]
 f_2  \\[6pt]
 f_3 
\end{array}\right)
=
\left(\begin{array}{c}
 \dfrac{F_x\xi}{W} \\[6pt]
\xi^2 \\[6pt]
\xi^2 s
\end{array}\right)
\ ,\qquad
\left(\begin{array}{c}
\dot x_1  \\[6pt]
\dot x_2 \\[6pt]
\dot x_3
\end{array}\right)
=
\left(\begin{array}{c}
\frac{{\rm d}\zeta }{{\rm d}\tilde t} \\[6pt]
\frac{{\rm d}\lambda }{{\rm d}\tilde t} \\[6pt]
\frac{{\rm d}\eta }{{\rm d}\tilde t}
\end{array}\right)
\ .
\end{equation}
The generalised force and generalised velocity are linearly related:
\begin{equation}
\label{eq: 2.3}
\boldsymbol{f}={\bf B}(\eta)\dot{\boldsymbol{x}}\,
\end{equation}
where $\boldsymbol{f} = (f_1\ f_2\ f_3)^T$,  
$\dot{\boldsymbol{x}} = (\dot x_1\ \dot x_2\ \dot x_3)^T$, and 
the components of the $3\times 3$ matrix ${\bf B}$ are given in Appendix A  
(they were derived in \cite{Wil+23}, where it was shown 
that they are functions of $\eta$, but are independent of the other coordinates). 
Linear equations of motion in the form of (\ref{eq: 2.3}) 
arise frequently in creeping flow problems. The matrix ${\bf B}$ is 
often referred to as the \emph{resistance matrix}~\cite{Hap+83}, and correspondingly 
we might refer to its inverse ${\bf C}={\bf B}^{-1}$ as a \emph{compliance matrix}.
The coefficients of ${\bf C}$ are also listed in  Appendix A. 

In the slider bearing problem, the independent variables are a mixture of forces and velocities.
Specifically, we fix two force components, namely the normalised vertical force $f_2=\xi^2$ 
and its moment $f_3=\xi^2 s$, and we also specify one velocity, namely the speed of the 
slider, $v$, by the dimensionless velocity 
\begin{equation}
\label{eq: 2.4}
\tilde v=\frac{\mu}{W}v
\ .
\end{equation}
The equation of motion should then yield the velocities $\dot x_2={\rm d}\lambda/{\rm d}\tilde t$  
and $\dot x_3={\rm d}\eta/{\rm d}\tilde t$. We should also expect to be able to calculate 
the dimensionless drag force, $f_1=F_x/W\xi$. 

Therefore, the problem of determining the equation of motion has the following 
structure. We have a matrix ${\bf C}$ which takes us from 
$\mbox{\boldmath$x$}=(x_1,x_2,x_3)$ to  $\mbox{\boldmath$y$}=(y_1,y_2,y_3)$, i.e., 
$\mbox{\boldmath$y$}={\bf C}\mbox{\boldmath$x$}$. But instead of knowing 
$\mbox{\boldmath$x$}$ and wanting to determine $\mbox{\boldmath$y$}$, we 
actually know $\mbox{\boldmath$u$}=(y_1,x_2,x_3)$ 
and want to determine $\mbox{\boldmath$v$}=(x_1,y_2,y_3)$ by writing 
$\mbox{\boldmath$v$}={\bf D}\mbox{\boldmath$u$}$, where 
${\bf D}$ is a $3\times 3$ matrix.

By means of a direct and mechanical calculation, the elements of ${\bf D}$ are found to be
\begin{equation}
\label{eq: 2.5}
{\bf D}=\frac{1}{C_{11}}\left(\begin{array}{ccc}
1 &  -C_{12} & -C_{13} \cr
C_{21}& C_{11}C_{22}-C_{21}C_{12}& C_{11}C_{23}-C_{21}C_{13}\cr
C_{31}& C_{11}C_{32}-C_{31}C_{12}& C_{11}C_{33}-C_{31}C_{13}\cr
\end{array}\right)
\ .
\end{equation}
(The notion of the \emph{Schur complement} (\cite{Zha05}) helps to simplify
and to generalise this calculation.) 
This re-arrangement of a linear equation of motion involves
exchanging the variables $x_1$ and $y_1$ between the set of 
dependent and  independent variables. It should find applications 
in other areas where there are linear equations of motion. We are 
not aware that the structure has been given a name. We propose 
that it could be referred to as a \emph{transversion} of the equations
of motion. 
 
We then have the equation of motion in the form
\begin{equation}
\label{eq: 2.6}
\left(\begin{array}{c}
f_1 \\[6pt]
\dot x_2 \\[6pt]
\dot x_3
\end{array}\right)
=
\left(\begin{array}{c}
\frac{F_x\xi}{W} \\[6pt]
\frac{{\rm d}\lambda }{{\rm d}\tilde t} \\[6pt]
\frac{{\rm d}\eta }{{\rm d}\tilde t}
\end{array}\right)
={\bf D}(\eta)
\left(\begin{array}{c}
\dot x_1 \\[6pt]
f_2 \\[6pt]
f_3
\end{array}\right)
={\bf D}(\eta)
\left(\begin{array}{c}
\tilde v \\[6pt]
\xi^2 \\[6pt]
\xi^2 s
\end{array}\right)
\ .
\end{equation}
The second and third rows yield two coupled equations of motion for 
$\lambda=\ln \xi$ and $\eta$:
\begin{eqnarray}
\label{eq: 2.7}
\frac{{\rm d}\lambda }{{\rm d}\tilde t}&=&
D_{21}(\eta)\tilde v+\xi^2\left[D_{22}(\eta)+sD_{23}(\eta)\right]\,
\nonumber \\
\frac{{\rm d}\eta }{{\rm d}\tilde t}&=&D_{31}(\eta)\tilde v+\xi^2\left[D_{32}(\eta)+sD_{33}(\eta)\right]
\ .
\end{eqnarray}
 Given the solutions $\lambda(\tilde t)$ and $\eta(\tilde t)$ we can then determine the 
 horizontal force:
 \begin{equation}
 \label{eq: 2.8}
 F_x=\frac{W}{\xi}\left(D_{11}(\eta)\tilde v+\xi^2\left[D_{12}(\eta)+sD_{13}(\eta)\right]\right)
 \ .
 \end{equation}
The full set of coefficients $D_{ij}$ are listed in Appendix A.
Some of these coefficients, in particular $D_{21}$ and $D_{31}$, 
are given exactly by very simple expressions.
The expressions for the coefficients occurring in the equations of motion 
for $\eta $ and $\lambda$ are:
\begin{eqnarray}
\label{eq: 2.9}
D_{21}=\frac{\eta}{2},&\qquad&D_{22}+sD_{23}=-16+30s+O(\eta),
\nonumber \\
D_{31}=-\frac{\eta^2}{2},&\qquad&D_{32}+sD_{33}=30-60s+O(\eta)
.
\end{eqnarray}
The equations of motion (\ref{eq: 2.7}) do not appear to allow an exact and general 
analytical solution, but in the following sections we address various cases.

\section{Steady motion}
\label{sec: 3}

\begin{figure*}
\centering
\includegraphics[width=0.45\textwidth]{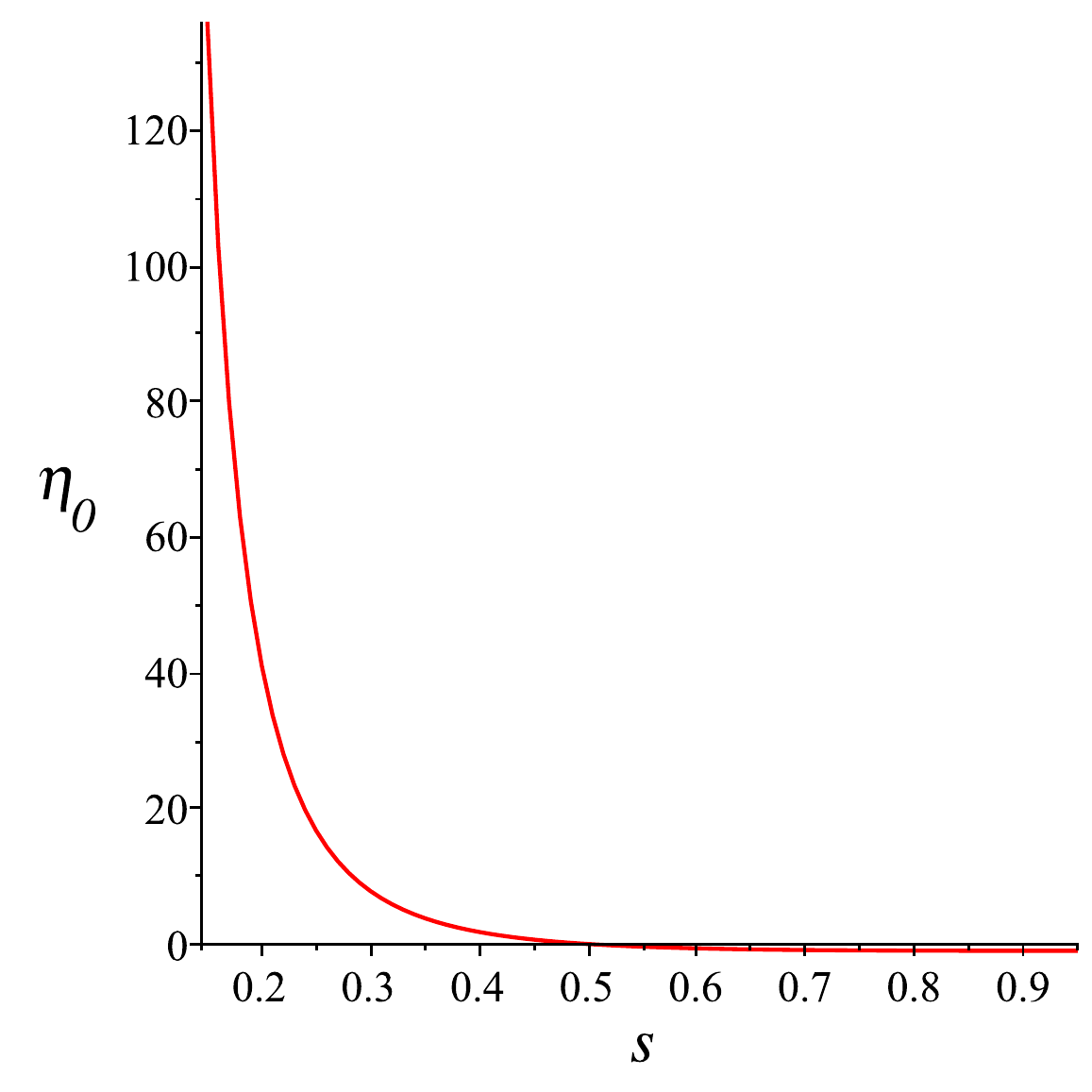}
\includegraphics[width=0.45\textwidth]{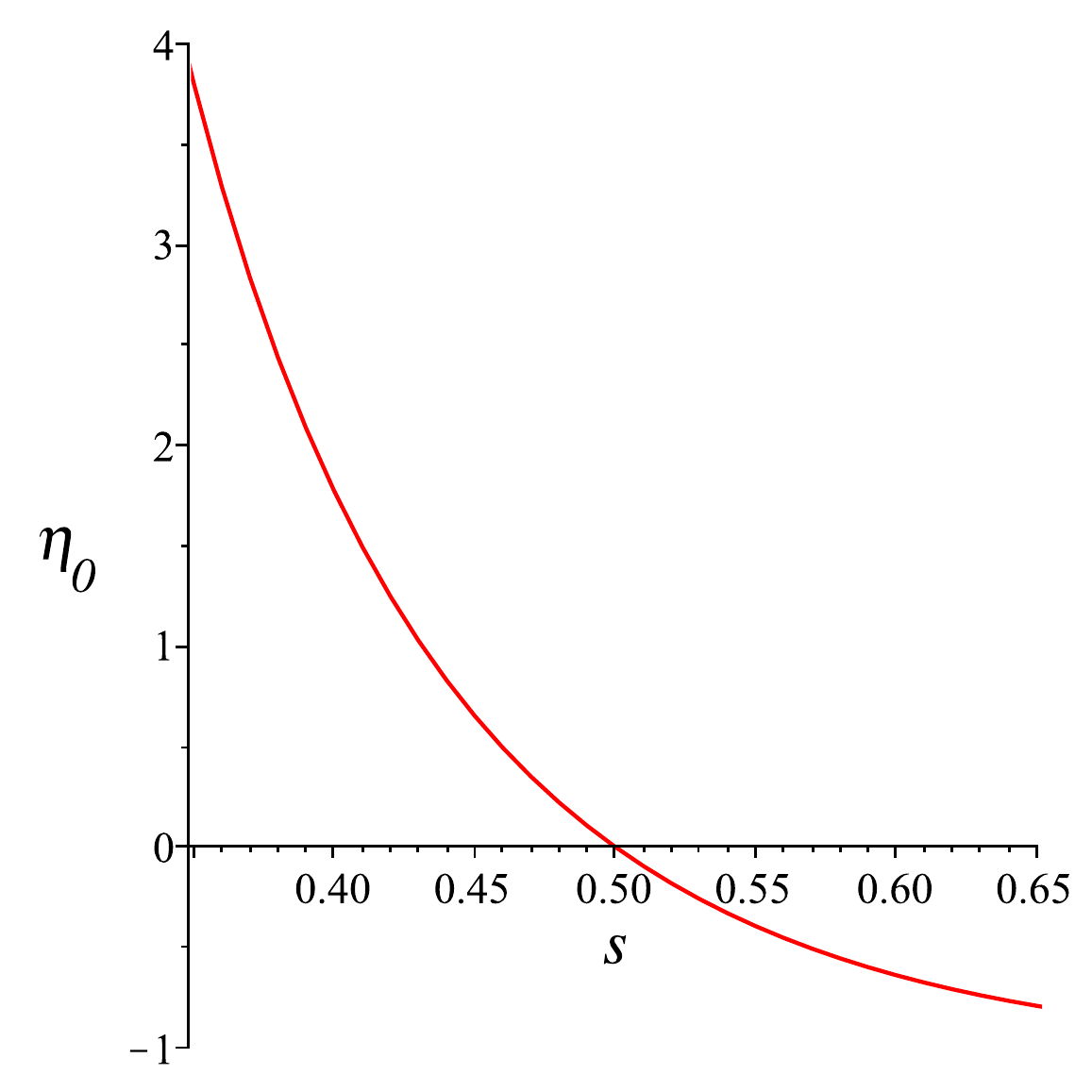}
\caption{
Plots of the steady-state value of $\eta$ as a function of  the pivot position, 
$s$. The right hand plot shows an expanded view of the area of particular interest.  
}
\label{fig: 3.1}
\end{figure*}

We seek a solution where the value of $\tilde v$ is given, and in which 
both $\eta$ and $\lambda$ are time-independent. Comparing the two 
equations in (\ref{eq: 2.7}), the condition for a fixed point is
 \begin{equation}
 \label{eq: 3.1}
 \frac{\tilde v}{\xi^2}=h_1(\eta)=h_2(\eta),
 \end{equation}
where
\begin{equation}
\label{eq: 3.2}
h_1(\eta)=-\frac{D_{22}(\eta)+sD_{23}(\eta)}{D_{21}(\eta)},\ \ \ 
h_2(\eta)=-\frac{D_{32}(\eta)+sD_{33}(\eta)}{D_{31}(\eta)}
\ .
\end{equation}
The fixed point value ,$\eta_0$, is therefore independent of $\tilde v$, and 
satisfies
\begin{equation}
\label{eq: 3.3}
h_1(\eta_0)=h_2(\eta_0)
\ .
\end{equation}
Equation (\ref{eq: 3.3}) shows that $\eta_0$ is a function of $s$, 
and solving this numerically yields the relationship illustrated in Fig.~\ref{fig: 3.1}.
Having determined $\eta_0$, the fixed point value of $\xi$ satisfies
\begin{equation}
\label{eq: 3.4}
\xi_0=\sqrt{\frac{\mu v}{W h_1(\eta_0)}}
\ .
\end{equation}
The horizontal force in the steady state is
\begin{equation}
\label{eq: 3.5}
F_x=W\xi_0\left[D_{11}(\eta_0)h_1(\eta_0)+D_{12}(\eta_0)+sD_{13}(\eta_0)\right]
\ .
\end{equation}
Note that $F_x\propto \tilde v^{1/2}$, which is different from the usual situation 
in viscous flow, where force is proportional to velocity. This unusual scaling arises 
because the angle and the gap of the slider change in order to support the load.

We can also analyse the steady-state solution by starting from equation equation (\ref{eq: 2.3}), 
and considering a solution with $\dot \lambda=\dot \eta=0$, by setting 
$\dot{\mbox{\boldmath$x$}}=(\tilde v,0,0)$, giving
\begin{subequations}
\label{eq: 3.6}
\begin{align}
F_x&=\frac{\mu}{\xi}B_{11}(\eta_0) v\\
W&=\frac{\mu}{\xi^2}B_{21}(\eta_0) v\\
WsL&=\frac{\mu L}{\xi^2}B_{31}(\eta_0) v
\end{align}
\end{subequations}
(these equations also correspond with equations (3.6) of \cite{Wil+23}).
For given $s$, these three equations are to be solved for $F_x$, $\eta$ and $\xi$.
Given the solution for $F_x$, the bearing performance is optimised by minimising 
the ratio $F_x/W$. Dividing the third equation of (\ref{eq: 3.6}) by the second gives
\begin{equation}
\label{eq: 3.7}
s(\eta_0)=\frac{B_{31}(\eta_0)}{B_{21}(\eta_0)}
={\frac { \left( -4\,{\eta_0}-6 \right) \ln  \left( {\eta_0}+1
 \right) +{{\eta_0}}^{2}+6\,{\eta_0}}{2{\eta_0}\, \left(  \left( {
\eta_0}+2 \right) \ln  \left( {\eta_0}+1 \right) -2\,{\eta_0}
 \right) }}
\end{equation}
which tells us that a fixed point of $\eta$ maps to a position along the bearing, 
which is independent of the load on the bearing and the horizontal velocity.  
(Here we used expressions for the coefficients $B_{ij}(\eta)$ given 
in Appendix A). Solving equation (\ref{eq: 3.7}) numerically again 
gives the steady-state value of $\eta$ as a function of $s$ illustrated 
in figure \ref{fig: 3.1}, as expected. 

\begin{figure}
\centering
\includegraphics[width=0.45\textwidth]{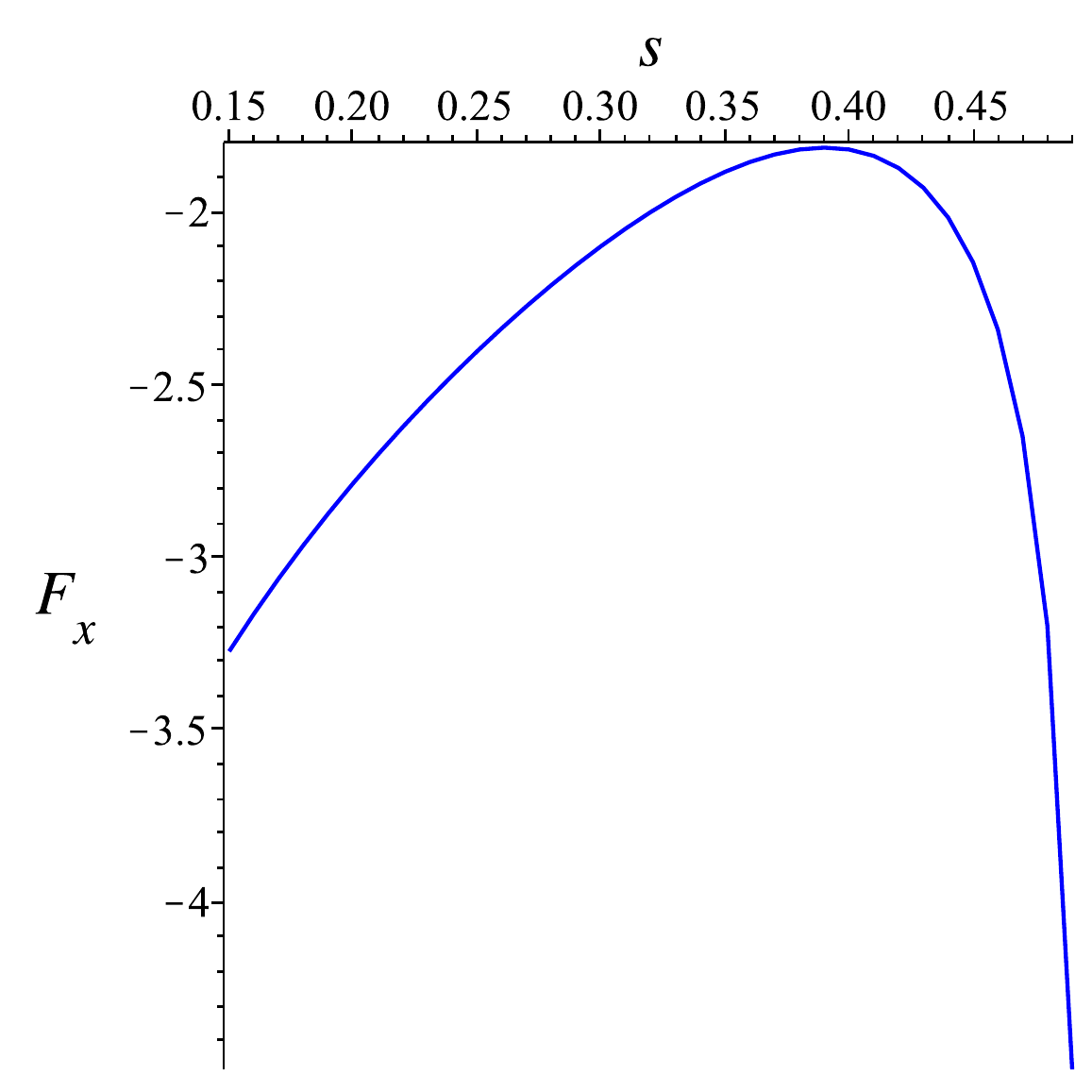}
\caption{
Plot of $F_x$ as a function of of the pivot position, $s$. 
}
\label{fig: 3.2}
\end{figure}

\begin{figure*}
\centering
\includegraphics[width=0.45\textwidth]{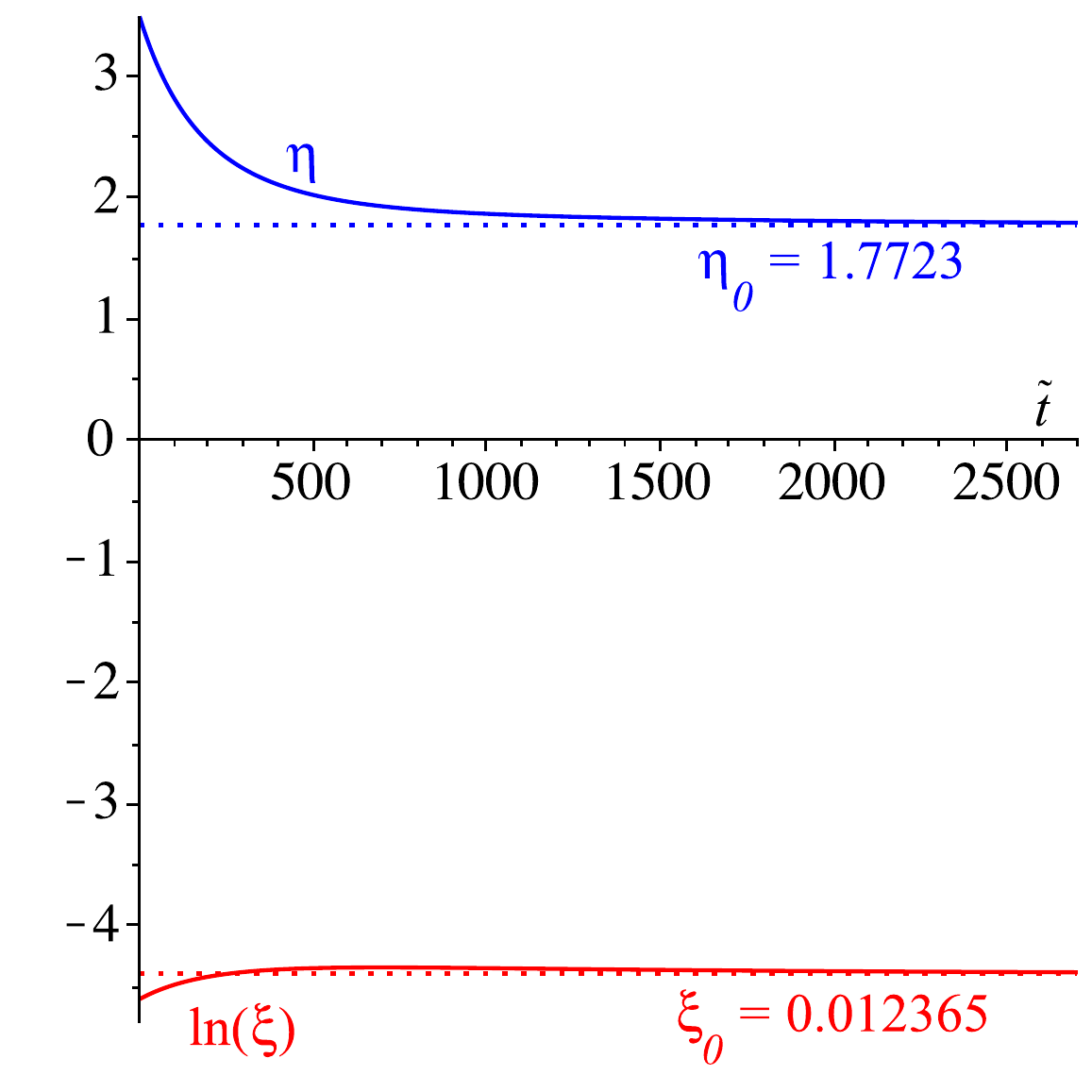}
\includegraphics[width=0.45\textwidth]{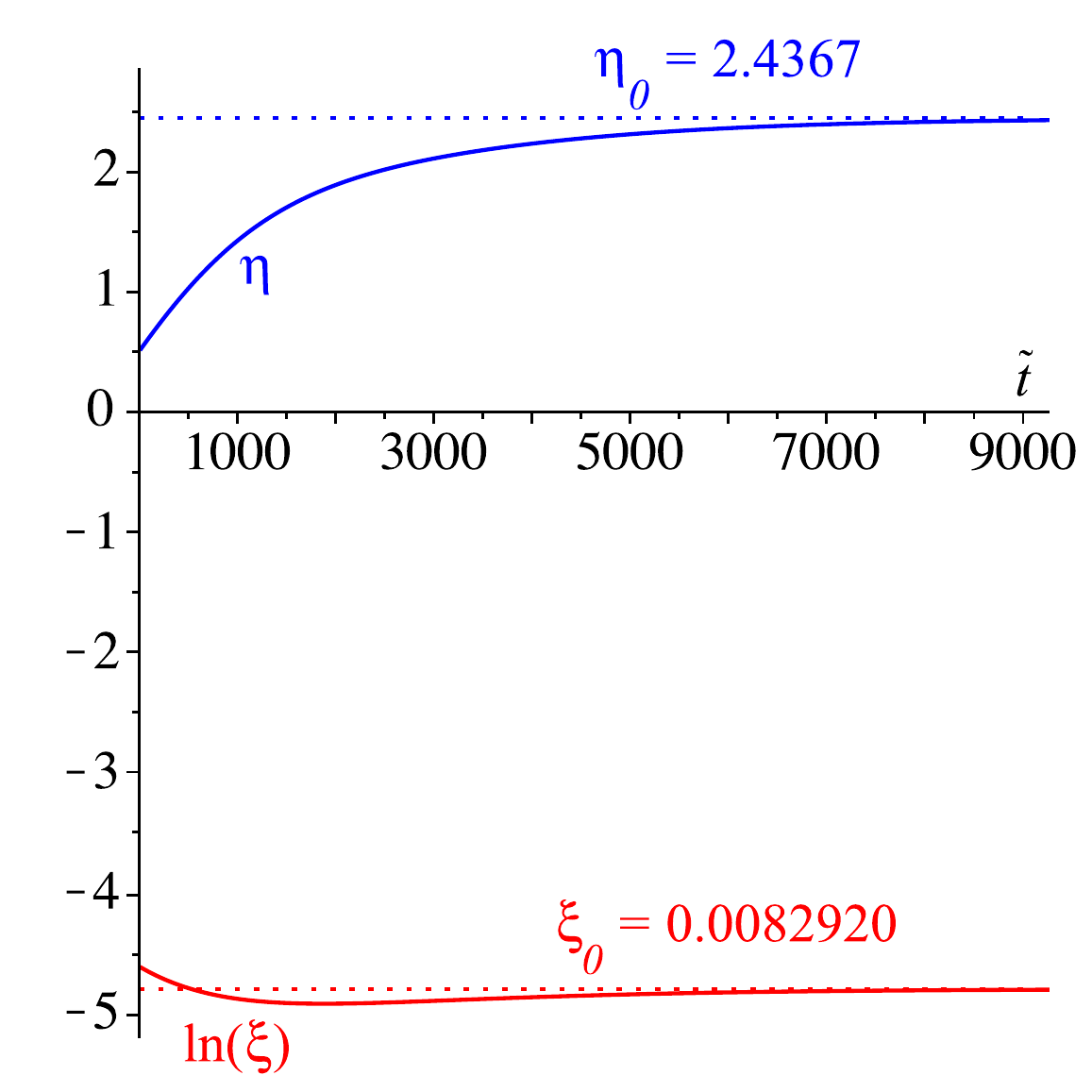}
\caption{
Plots illustrating relaxation of a pivoted bearing towards its equilibrium 
configuration when moving with a constant speed. The left panel shows the pivot at $s=0.4$ and 
speed $\tilde v = 0.001$ with initial conditions $\eta_{\rm in}=3.5$, $\xi_{\rm in}=0.01$. 
The right panel shows the pivot at $s=0.38$ and speed $\tilde v = 0.0005$ with initial 
conditions $\eta_{\rm in}=0.5$, $\xi_{in}=0.01$.
}
\label{fig: 3.3}
\end{figure*}

The first two equations of (\ref{eq: 3.6}) lead to
\begin{equation}
\label{eq: 3.8}
F_x=\frac{B_{11}(\eta_0(s))}{\sqrt{B_{21}(\eta_0(s))}}\sqrt{\mu vW}
\ .
\end{equation}
Figure \ref{fig: 3.2} shows the resistive force as a function of $s$, when $\mu vW=1$.
The optimum pivot, which minimises the resistive force, is at 
position $s^\ast=0.3904\ldots$, which is 
independent of $\tilde v$, and corresponds to $\eta=\eta_0=2.0713\ldots$.
The resistive force for this optimal pivot position is $F_x=-1.8125\ldots \times \sqrt{\mu v W}$.
We also used Maple to solve equations (\ref{eq: 2.6}) numerically. 
Figure \ref{fig: 3.3} illustrates convergence to 
the steady state solution predicted by (\ref{eq: 3.3}) and (\ref{eq: 3.4}).

\section{Stability of steady flow}
\label{sec: 4}

Figure \ref{fig: 3.3} shows convergence towards the steady-state 
solutions. In this section, we consider whether the steady-state solutions are stable 
in the general case. 

\begin{figure*}
\centering
\includegraphics[width=0.45\textwidth]{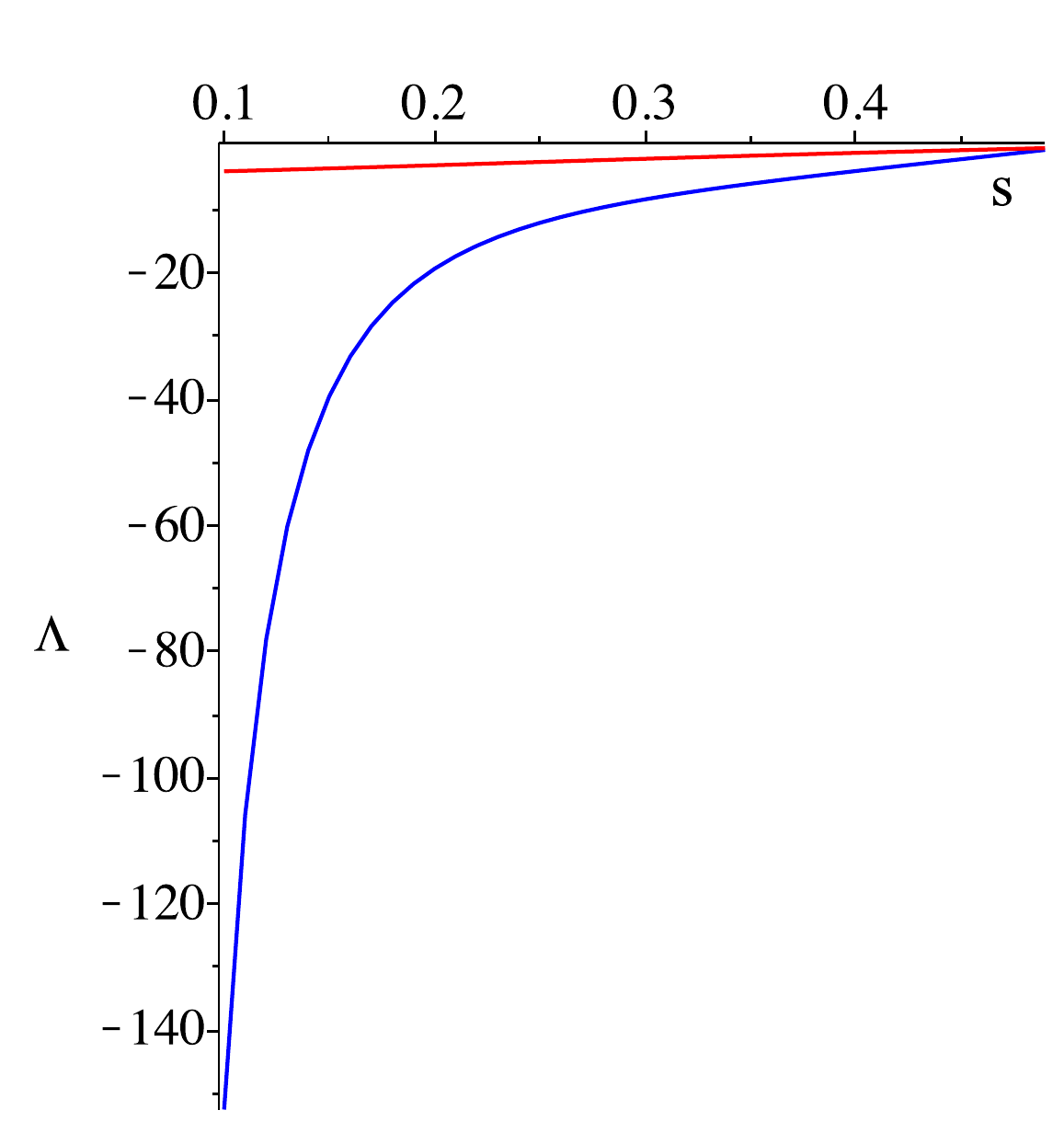}
\includegraphics[width=0.45\textwidth]{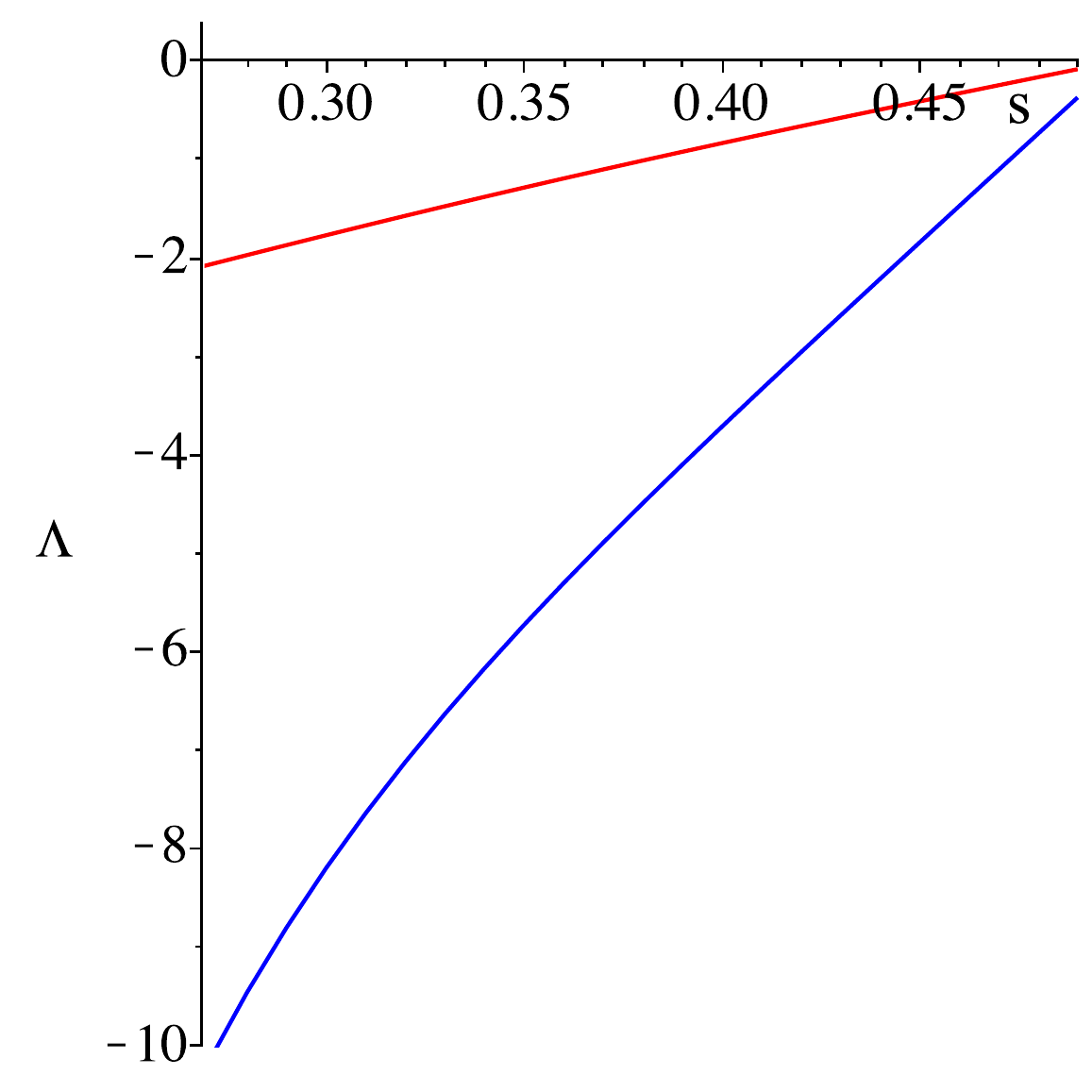}
\caption{
Plot of the eigenvalues of the matrix ${\bf M}$ as a function of the pivot point position 
$s$ (which defines the steady-state aspect parameter $\eta_0$). 
These define the stability of the steady-state
solution. Since they are both negative the steady-state solutions are locally stable. 
Values calculated at 0.01 intervals from $s=0.1$ to $s=0.49$. 
Right hand panel shows expanded view of the region where both eigenvalues $> -10$.
}
\label{fig: 4.1}
\end{figure*}

\begin{figure*}
\centering
\includegraphics[width=0.45\textwidth]{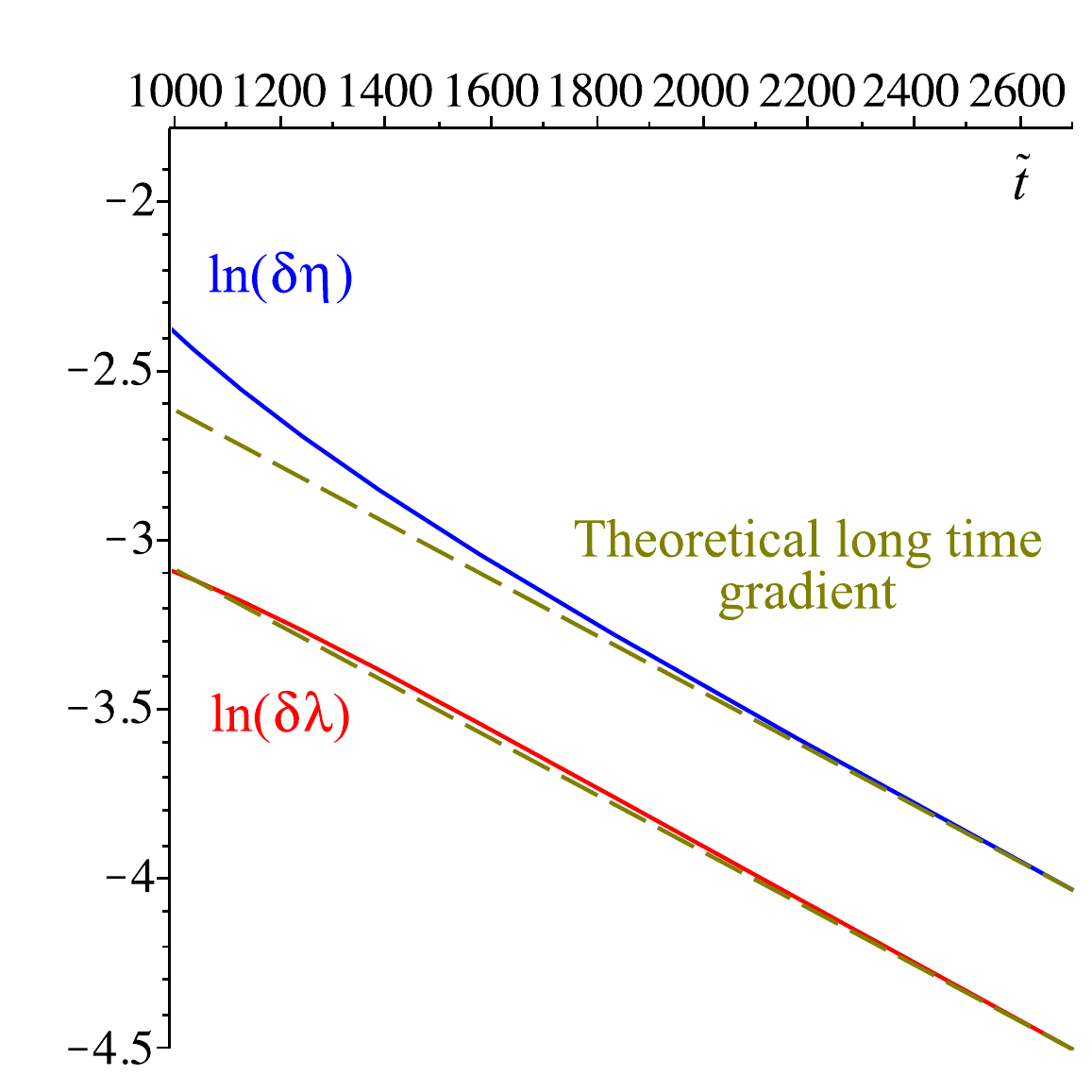}
\includegraphics[width=0.45\textwidth]{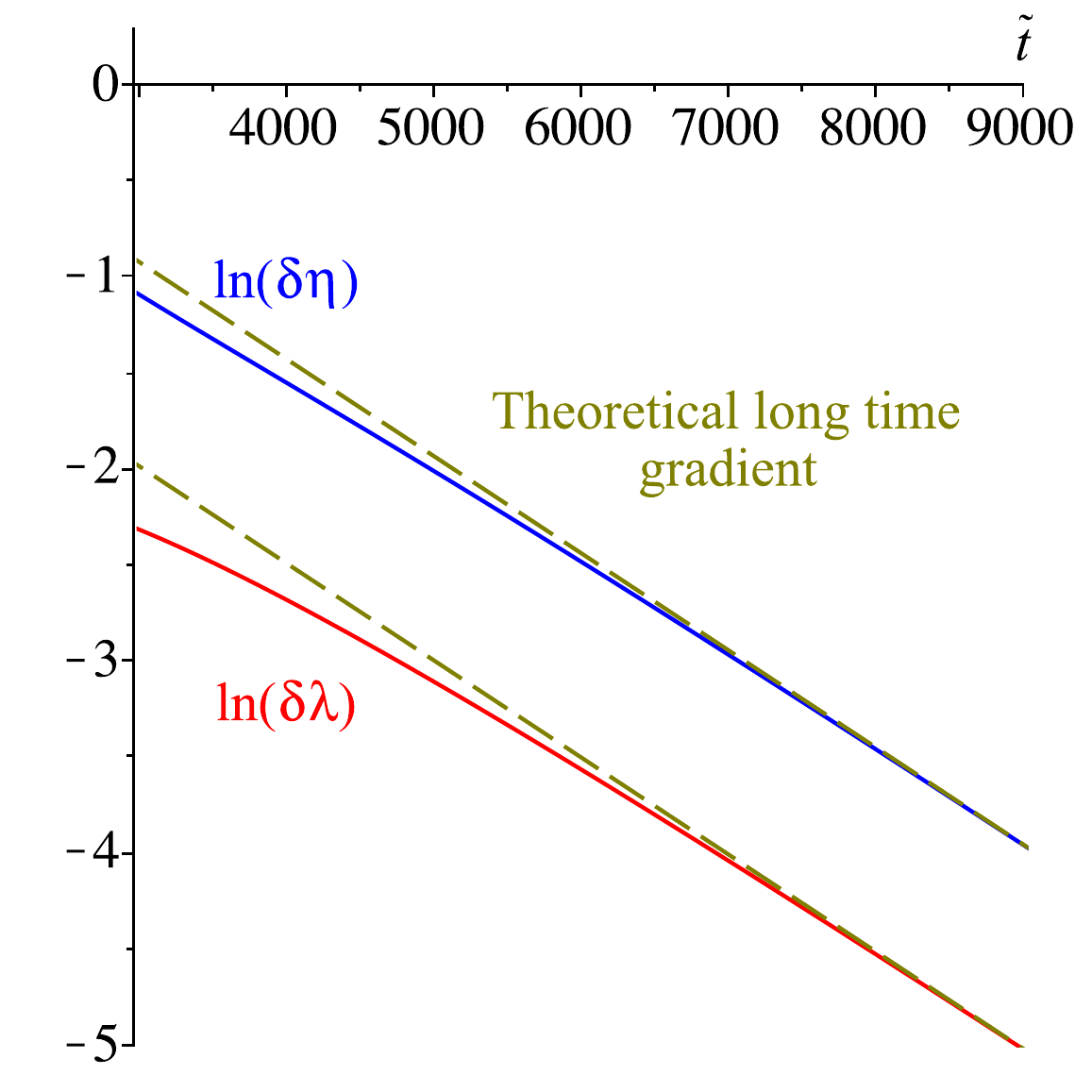}
\caption{
Plots of $\ln (\delta \eta)$ and $\ln (\delta \lambda)$ as they approach their 
equilibrium values for the situations shown in figure \ref{fig: 3.3}, with 
superimposed (olive) lines showing their theoretical gradients 
(which are the bearing speed multiplied by the minimum 
magnitude eigenvalue of the matrix ${\bf M}$).
}
\label{fig: 4.2}
\end{figure*}

Using (\ref{eq: 2.9}), the equations of motion (\ref{eq: 2.7}) can be written in the form
\begin{eqnarray}
\label{eq: 4.1}
\frac{{\rm d}\lambda}{{\rm d}\tilde t}&=&\frac{\eta \tilde v}{2}+\exp(2\lambda)A(\eta),
\nonumber \\
\frac{{\rm d}\eta}{{\rm d}\tilde t}&=&-\frac{\eta^2\tilde v}{2}+\exp(2\lambda)B(\eta).
\end{eqnarray}
The steady state conditions give the relations
\begin{equation}
\label{eq: 4.2}
A(\eta_0)\eta_0=-B(\eta_0)
\ ,\ \ \ 
\xi_0^2=-\eta_0\tilde v/2A(\eta_0)
\ .
\end{equation}
Considering small deviations from the steady-state 
$\eta(\tilde t)=\eta_0+\delta \eta(\tilde t)$ and $\lambda(\tilde t)=\lambda_0+\delta \lambda(\tilde t)$, 
we have
\begin{equation}
\label{eq: 4.3}
\left(\begin{array}{c}\frac{{\rm d}\delta \lambda}{{\rm d}\tilde t}\\[6pt]
\frac{{\rm d}\delta \eta}{{\rm d}\tilde t}\end{array}\right)=\tilde v {\bf M}(\eta_0)
\left(\begin{array}{c}\delta \lambda \\[6pt]
\delta \eta\end{array}\right)
\end{equation}
where 
\begin{equation}
\label{eq: 4.4}
{\bf M}(\eta)=
\left(\begin{array}{cc}
-\eta&\frac{1}{2}\left(1-\frac{\eta A'(\eta)}{A(\eta)}\right)\cr
-\frac{\eta B(\eta)}{A(\eta)}&-\eta\left(1+\frac{B'(\eta)}{2A(\eta)}\right)\cr
\end{array}\right)
\ .
\end{equation}
The solution of equation (\ref{eq: 4.3}) may be expressed in terms of the 
eigenvectors, $\mbox{\boldmath$e$}_i$ and eigenvalues, $\Lambda_i$, of the matrix ${\bf M}$.
In particular, the steady-state solutions are locally stable if the eigenvalues $\Lambda_i$ 
are both negative. These eigenvalues are plotted, as a function of $s$, in Fig.~\ref{fig: 4.1}.  
The fact that the eigenvalues are always negative implies that the 
steady-state solutions, investigated in \cite{Mic50}, are in fact 
always stable. The timescale for convergence to the equilibrium values, 
$\eta_0$, $\xi_0$, is determined by the inverse of the eigenvalue which is 
smaller in magnitude. The eigenvalues are proportional to the elements of the matrix ${\bf M}$, 
and therefore to $\tilde v$. This implies that the relaxation in the dimensionless time $\tilde t$ 
is of order $1/\tilde v$. In terms of dimensioned  quantities, the relaxation time is 
therefore of order $L/v$, which is the timescale for the lubricant film to pass under the bearing, 
as might be expected on physical grounds. 
Relaxation toward the steady-state is illustrated in Fig.~\ref{fig: 4.2}, for the cases
which were treated in Fig.~\ref{fig: 3.3}.
In section \ref{sec: 6} we show that the 
equilibration time can be much greater than $L/v$ when the final velocity is sufficiently 
large that this perturbative approach is not applicable.

\section{Sinking of stationary slider}
\label{sec: 5}

In this section we consider the settling of the bearing when it stops moving, so 
that $\tilde v=0$. Under these conditions, equations (\ref{eq: 2.7}) give
\begin{eqnarray}
\label{eq: 5.1}
\frac{{\rm d}\xi }{{\rm d}\tilde t}&=&\xi^3\left[D_{22}(\eta)+sD_{23}(\eta)\right]
\equiv \xi^3 A(\eta),
\nonumber \\
\frac{{\rm d}\eta}{{\rm d}\tilde t}&=&\xi^2\left[D_{32}(\eta)+sD_{33}(\eta)\right]
\equiv \xi^2 B(\eta).
\end{eqnarray}
If there is a solution of the equation 
\begin{equation}
\label{eq: 5.2}
B(\eta_-)=0,
\end{equation}
then there is a solution for which the bearing sinks with its aspect 
parameter $\eta$  equal to $\eta_-$. 
In the following we obtain $\xi(\tilde t)$ for this solution, and 
investigate the stability of $\eta$ under a perturbation about
its fixed point, $\eta_-$.

If $\eta$ is constant, the first equation of (\ref{eq: 5.1}) gives
\begin{equation}
\label{eq: 5.3}
\xi^2 = \frac{\xi(0)^2}{1- 2\xi(0)^2 A(\eta_-) \tilde t},
\end{equation}
so $\xi\sim \tilde t^{-1/2}$ as $\tilde t\to \infty$. 
If $\delta \eta$ is a small perturbation about the fixed point $\eta_-$ satisfying
$B(\eta_-)=0$, then
\begin{equation}
\label{eq: 5.4}
\frac{{\rm d}\,\delta \eta}{{\rm d}\,\tilde t}\sim \xi^2 B'(\eta_-)\delta \eta
+2\xi B(\eta_-)\delta \xi 
= \xi^2 \delta \eta B'(\eta_-)
\ .
\end{equation}
Combining this with the first equation of (\ref{eq: 5.1}), we have
\begin{equation}
\label{eq: 5.5}
\frac{{\rm d}\,\delta \eta}{{\rm d}\xi}= \frac{\delta \eta}{\xi} \frac{B'(\eta_-)}{A(\eta_-)},
\end{equation}
so, 
\begin{equation}
\label{eq: 5.6}
\delta \eta (\tilde t)= \delta \eta (0) 
\left( \frac{\xi (\tilde t)}{\xi(0)}\right)^{\frac{B'(\eta_-)}{A(\eta_-)}},
\end{equation}
and as $\tilde t \to \infty$, $\xi (\tilde t)\to 0$, so $\delta \eta \to 0$ and the solution is stable if 
$B'(\eta_-)/A(\eta_-)>0$. 

\begin{figure}
\centering
\includegraphics[width=0.45\textwidth]{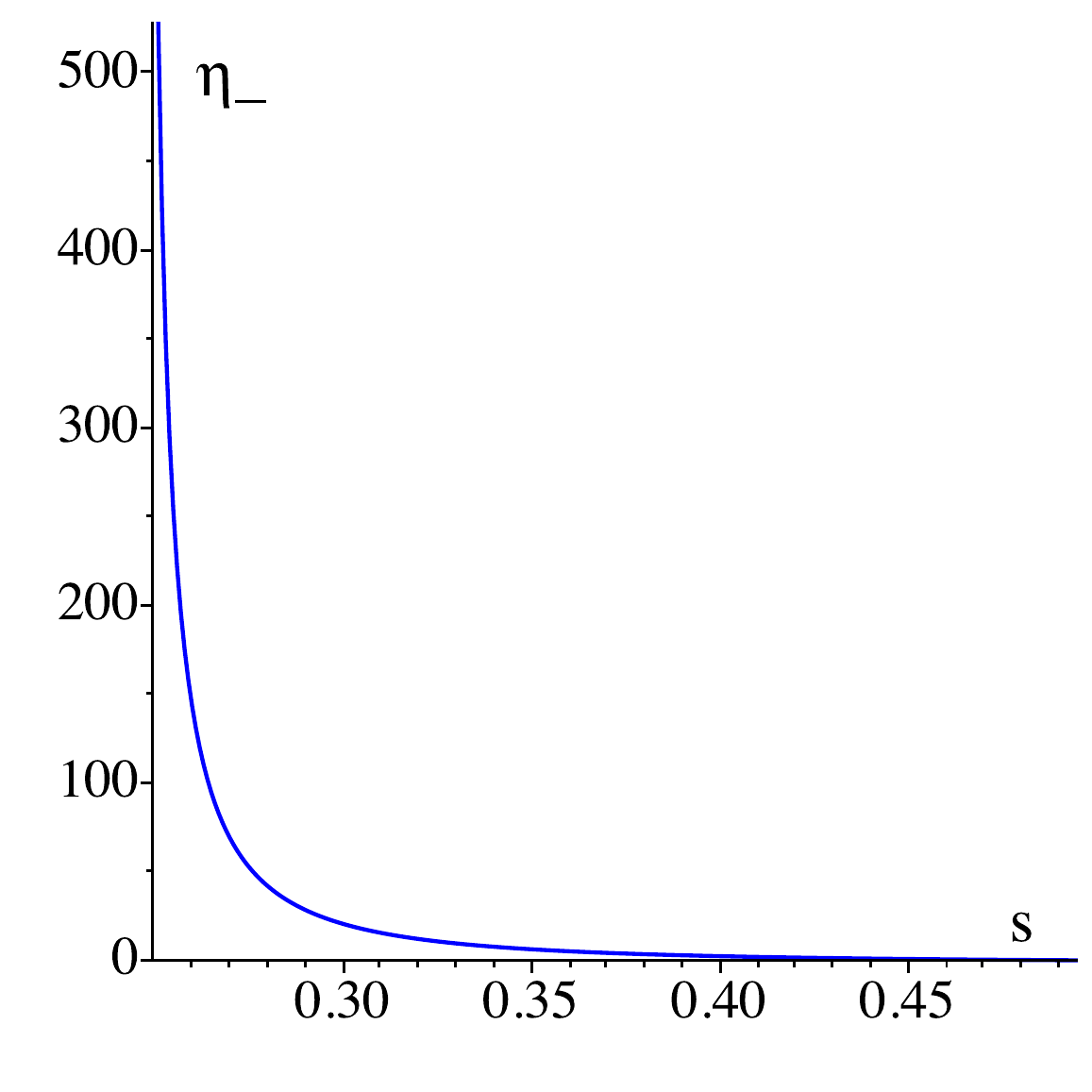}
\includegraphics[width=0.45\textwidth]{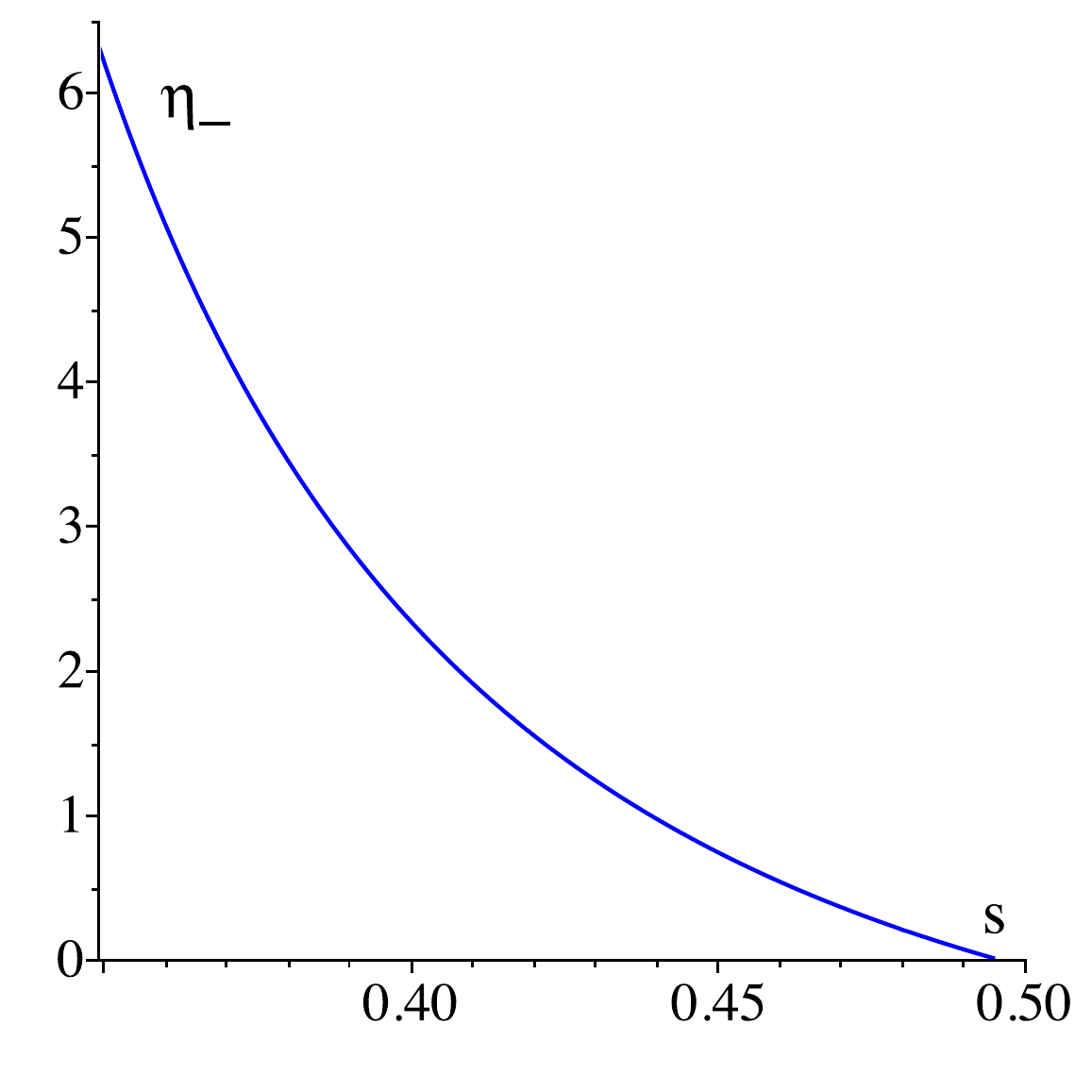}
\caption{
Plot of the fixed point aspect parameter, $\eta_-$, for sinking motion of a stationary 
slider, as a function of $s$. The right hand panel shows expanded view of the region where $s>0.35$.  
}
\label{fig: 5.1.1}
\end{figure}

Now let us consider whether a solution satisfying $B(\eta_-)=0$ exits, and whether it is stable.
Substituting for $D_{32}$ and $D_{33}$ (using expressions given in Appendix A), and defining 
$\psi=\ln(1+\eta)$, leads to an expression of the form
\begin{equation}
\label{eq: 5.7}
B(\eta)=\frac{\eta^4[2 (1+\eta)\psi + \eta^2 (1 - 4s) -2\eta]}
{12(2\psi^2 \eta + \psi \eta^2 + 2\psi^2 + 2\psi \eta -4\eta^2)} \equiv \frac{\eta^4 C(\eta)}{ D(\eta)},
\end{equation}
which defines functions $C(\eta) $ and $D(\eta)$. Since $D(\eta)>0$ for $\eta>0$, the condition $B(\eta_-)=0$ 
is then solved by finding a solution to $C(\eta_-)=0$, or else by $\eta_-=0$.
Considering the first of these possibilities, if $C(\eta_-)=0$ since we have $C(0)=0$ 
and $C'(\eta) = 2 \ln(1+\eta) + 2\eta (1 - 4s)$, then $C(\eta)$ cannot be $0$ for 
$\eta>0$ and $0 < s \leq \frac{1}{4}$, so there is no solution to the equation $B(\eta_-)=0$ 
for $s$ in this range. However, considering $C(\eta_-)=0$ for $\eta>0$ and 
$\frac{1}{4} < s < \frac{1}{2}$, here $C'(\eta)$ always has a single zero, corresponding 
to a maximum of $C(\eta)$, which is positioned at $\eta \to 0$ as $s \to \frac{1}{2}$ from 
below and at $\eta \to \infty$ as $s \to \frac{1}{4}$ from above. Hence there is an $\eta_-$ for 
which $C(\eta_-)=0$ which occurs to the right of the zero of $C'(\eta)$, i.e., $0<\eta_-<\infty$, 
with $\eta_- \to \infty$ as $s \to \frac{1}{4}$ from above. Figure \ref{fig: 5.1.1} illustrates this solution $\eta_-$ of (\ref{eq: 5.2}) as a function of $s$.

\begin{figure}
\centering
\includegraphics[width=0.4\textwidth]{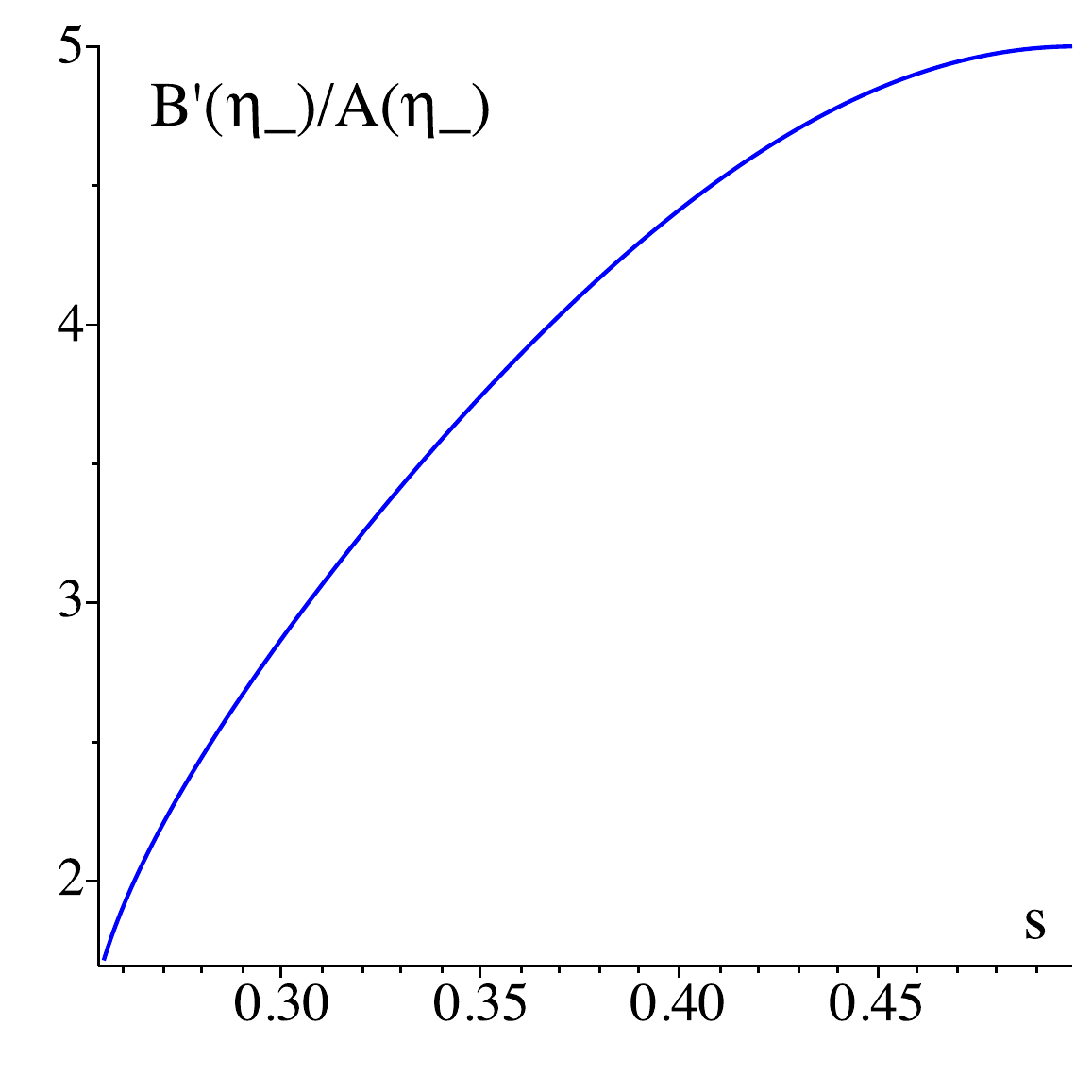}
\caption{\label{fig: 5.4}
Plot of $B'(\eta_-)/A(\eta_-)$ for pivot point values in the interval between $s=1/4$  and $s=1/2$. 
}
\end{figure}

\begin{figure*}
\centering
\includegraphics[width=0.45\textwidth]{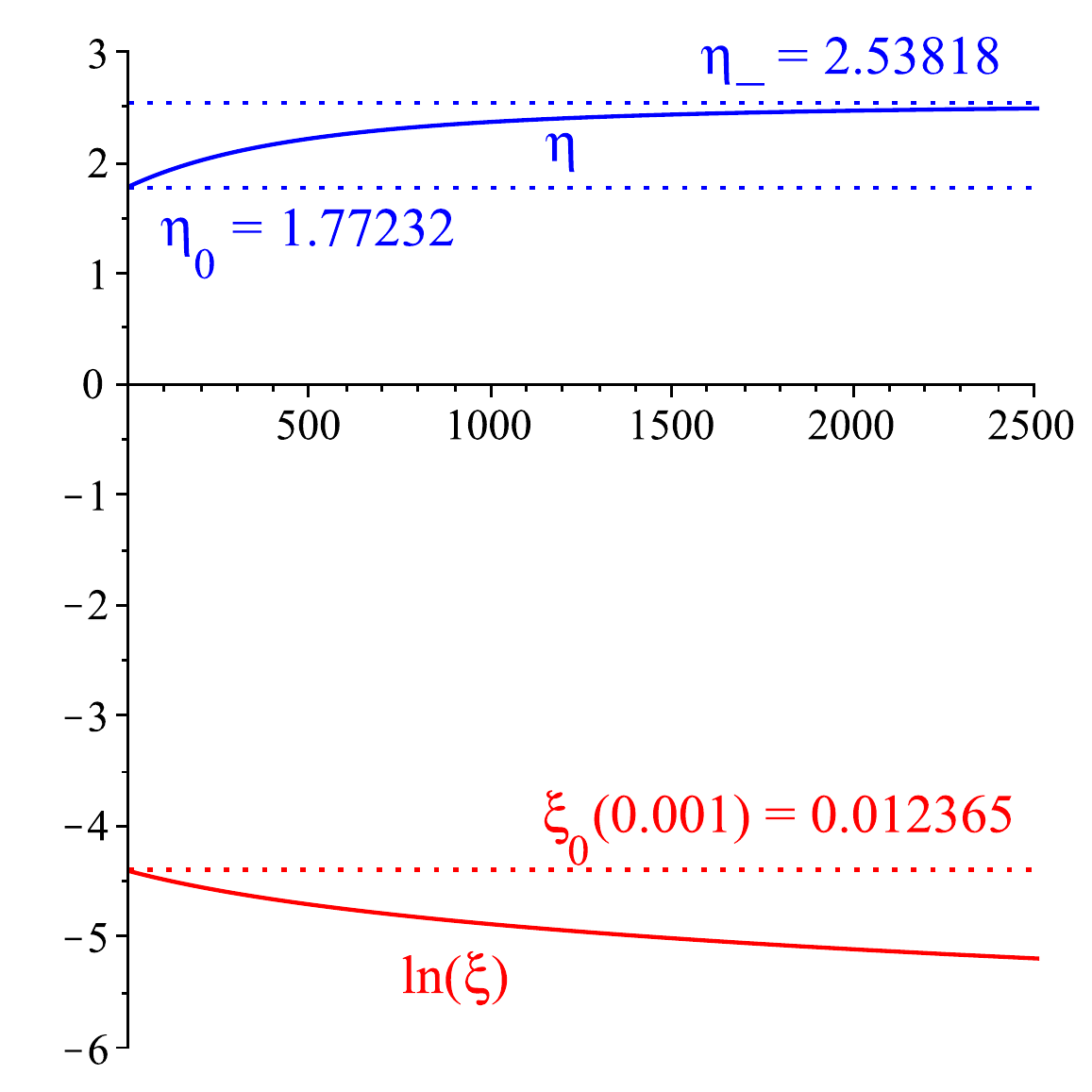}
\includegraphics[width=0.45\textwidth]{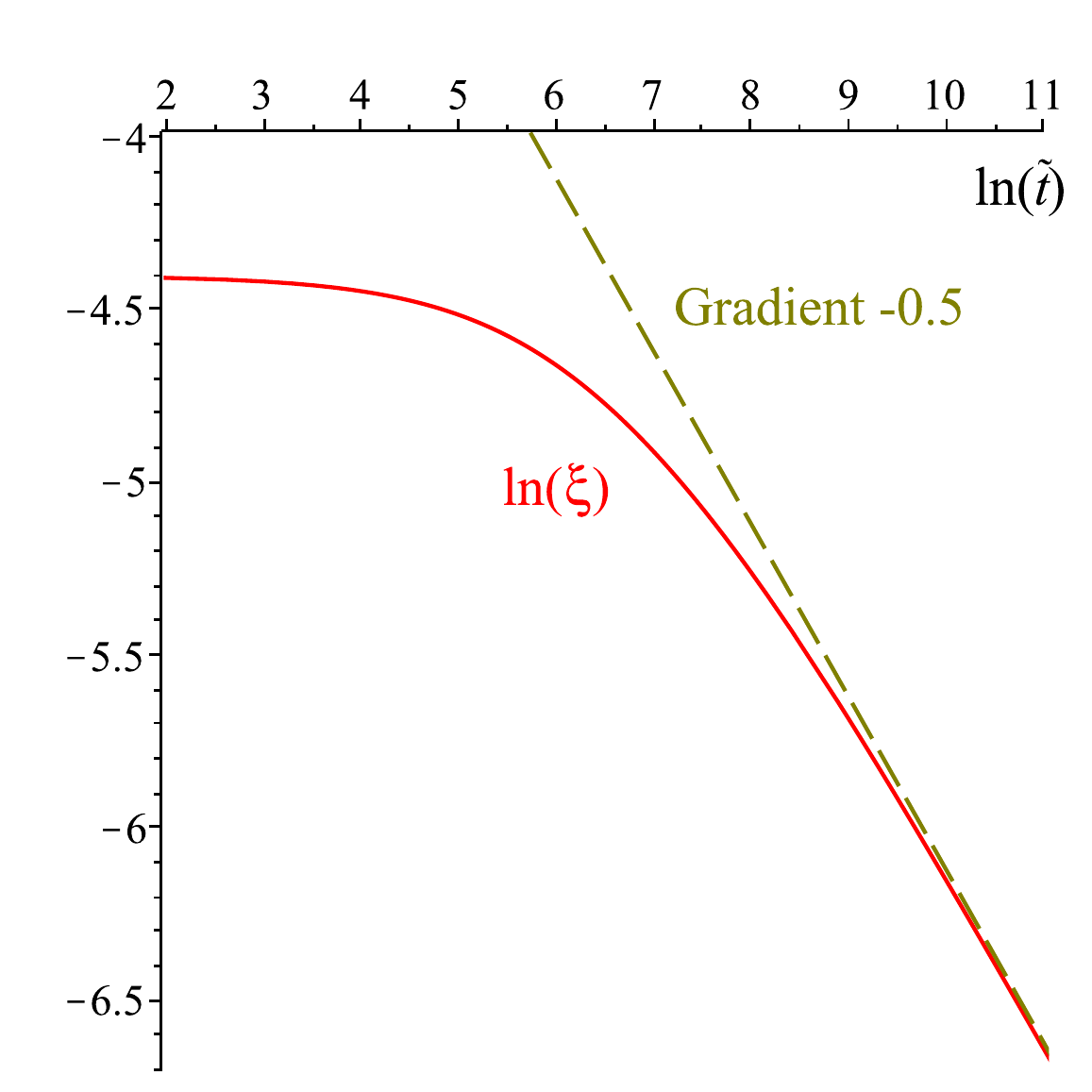}
\includegraphics[width=0.45\textwidth]{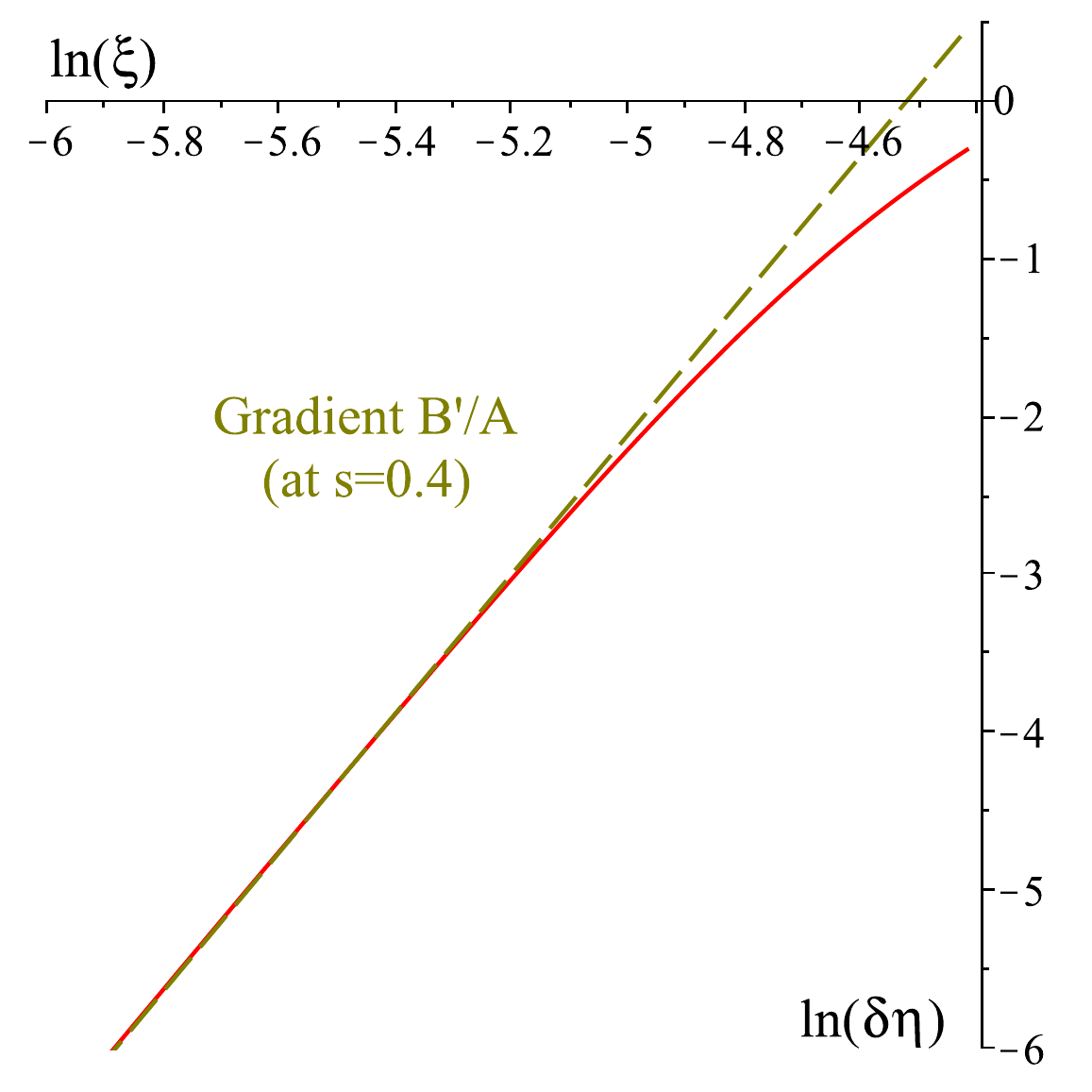}
\caption{Plots illustrating the motion of a bearing pivoted at $s=0.4$ 
starting from dynamic equilibrium, then settling after the speed is reduced to zero. 
Top left-hand panel: initial speed 
$v=0.001$ is suddenly reduced to zero. The value of $\eta$ changes from 
$\eta\approx\eta_0$ to $\eta\approx \eta_-$.
Top right-hand panel: illustrates that the settling is asymptotic to 
$\xi\sim \tilde t^{-1/2}$. Bottom panel illustrates that $\delta \eta$ is asymptotic to 
$\left[ \xi (\tilde t)\right]^{\left( \frac{B'(\eta_-)}{A(\eta_-)}\right)}$
}
\label{fig: 5.1}
\end{figure*}

Equation (\ref{eq: 5.7}) suggests that $\eta_-=0$ may be another solution of (\ref{eq: 5.2}), 
but $D(\eta)$ also vanishes as $\eta\to 0$. Application of L'Hospital's rule gives 
$B(0)=30(1 - 2s)$, so $\eta_-=0$ is only a solution for $s=\frac{1}{2}$, i.e., the pivot point is in 
the middle of the slider. This is coincident with the limit of the solution of $S(\eta)=0$ 
as $s\to 1/2$.
Equation (\ref{eq: 5.6}) shows that the solution with constant $\eta$  is stable 
provided $B'(\eta_-)/A(\eta_-)$ is positive.  In figure \ref{fig: 5.4} this quantity  
is plotted as a function of $s$, which defines $\eta_-$, confirming the solution is stable 
for $\frac{1}{4}<s<\frac{1}{2}$.  Figure \ref{fig: 5.1} illustrates examples of this type of motion.
For values of $s \leq \frac{1}{4}$ where there are no real solutions to equation 
(\ref{eq: 5.2}), we hypothesise that the bearing would make contact with the plane in finite time.  

\section{Response to large increase of velocity}
\label{sec: 6}

\begin{figure}
\centering
\includegraphics[width=0.65\textwidth]{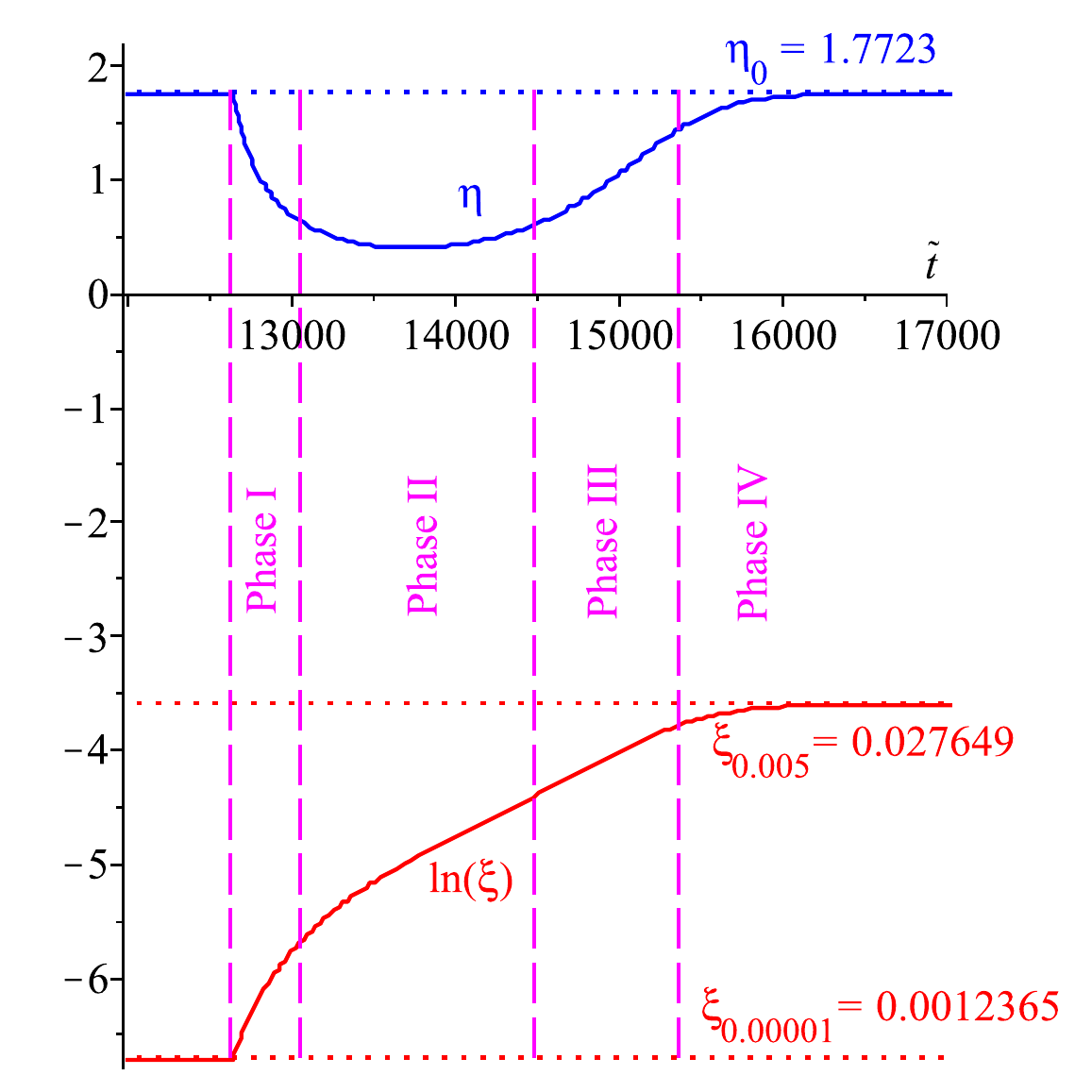}
\caption{
A bearing pivoted at $s=0.4$ is established in its equilibrium configuration 
when first moving with a constant speed of $0.00001$. Its speed is then
instantaneously increased by a factor of $500$ to $0.005$ and it relaxes
towards a new equilibrium with $\eta_0$ unchanged, but a 
different $\xi_0$ (shown as $\xi_{0.00001}$ and $\xi_{0.005}$).
The evolution, described by equations (\ref{eq: 4.1}), can be divided into four phases. 
In phase I, $\xi$ is sufficiently small that the terms in $\exp(2\lambda)$ can be 
neglected in both equations. In phase II, the two terms on the RHS of the second 
equation are comparable, but the term in the first equation proportional to 
$\exp(2\lambda)$ remains negligible. Later, in phase III, the term 
proportional to $\exp(2\lambda)$ dominates the RHS of the second equation, so 
that $\eta$ increases. Finally, in phase IV, a new equilibrium is established.
}
\label{fig: 6.1}
\end{figure}

Let us consider what happens when the coordinates of the slider are 
initially $\xi_{\rm in}$ and $\eta_{\rm in}$, and the velocity $\tilde v$ is suddenly 
increased to a \lq large' value, such that the steady state gap $\xi_0$ 
satisfies $\xi_0/\xi_{\rm in}\gg 1$. 
We argue that, when $\xi_0/\xi_{\rm in}\gg 1$, the value of $\eta$ falls to a 
very low value, $\underline \eta$, before increasing to $\eta_0$. 
This behaviour can be observed in Fig.~\ref{fig: 6.1}. Let us consider
why this happens, and how this mechanism leads to a strategy for analysing this transient. 

Consider the evolution of $\lambda(\tilde t)$ and $\eta(\tilde t)$ 
as described by equations (\ref{eq: 4.1}), illustrated in figure \ref{fig: 6.1}.
At the start of the transient, the smallness of $\xi_{\rm in}$ ensures that 
only the terms proportional to $\tilde v$ on the RHS of 
equations (\ref{eq: 4.1}) are significant. If the terms proportional to $\xi^2=\exp(2\lambda)$ are 
dropped, the resulting equations are analytically solvable. The solutions of these truncated 
equations are valid in the initial stage (\emph{phase I}) of the transient, implying that 
$\lambda$ increases and $\eta$ decreases, until we reach a stage, (\emph{phase II}), 
where the two terms on the RHS of the \emph{second} equation of (\ref{eq: 4.1}) are 
comparable. Note that, because the first equation of (\ref{eq: 4.1}) has a different 
dependence upon $\eta$, the term proportional to $\exp(2\lambda)$ remains 
negligible in the first equation of (\ref{eq: 4.1}) during phase II. 

The evolution of $\lambda$ and $\eta$ during these first two phases 
of the transient is, therefore, described by the following truncated approximation of (\ref{eq: 4.1}):
\begin{eqnarray}
\label{eq: 6.1}
\frac{{\rm d}\lambda }{{\rm d}\tilde t}&=&\frac{\tilde v}{2}\eta,
\\
\label{eq: 6.2}
\frac{{\rm d}\eta }{{\rm d}\tilde t}&=&-\frac{\tilde v}{2}\eta^2+\exp(2\lambda)
\left[(30-60s)+{\rm O}(\eta)\right].
\end{eqnarray}
The value of $\eta$ reaches its minimum $\underline \eta$ during phase II, 
at time $\underline {\tilde t}$. At this point, the two terms on the RHS of (\ref{eq: 6.2}) 
balance, while ${\rm d}\lambda/{\rm d}\tilde t$ is positive. As $\tilde t$ increases, 
we reach \emph{phase III} of the transient, where the term proportional to $\exp(2\lambda)$ 
on the RHS of (\ref{eq: 6.2}) is dominant, which results in $\eta$ increasing. 
In the final, \emph{phase IV}, of the transient, $\lambda$ becomes 
sufficiently large that the term which was dropped on the RHS of (\ref{eq: 6.1}), 
proportional to $\exp(2\lambda)$, becomes significant. In this final phase, the 
values of $\xi=\exp(\lambda)$ and $\eta$ approach the stable fixed point 
values, $\xi_0$ and $\eta_0$, which corresponds to 
the velocity $v$ and pivot position $s$.

We are primarily concerned with estimating the energy dissipation during the transient,
which is dominated by the initial phases, when $\xi$ is small.
In sub-section \ref{sec: 6.1} we consider the evolution of $\eta$ and $\lambda$ 
during phases I and II, and we make a simple estimate of the values of $\underline {\eta}$ and 
$\underline{\tilde t}$ based upon these solutions. We then (sub-section \ref{sec: 6.2}) 
use these approximate solutions 
for $\eta(\tilde t)$ and $\lambda(\tilde t)$ to estimate the excess energy 
dissipated.

\subsection{Initial phases of the transient}
\label{sec: 6.1} 

\begin{figure}
\centering
\includegraphics[width=0.99\textwidth]{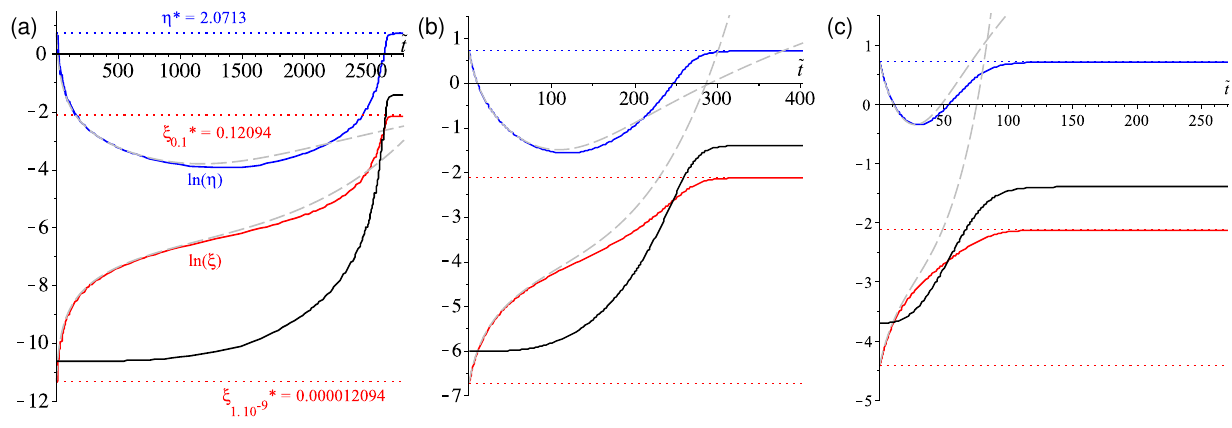}
\caption{Dynamics of a slider bearing pivoted at $s=0.3904$ with  speed increasing 
instantaneously by three different factors: ({\bf a}) a factor of $10^8$, 
from $0.000000001$ to $0.1$ ,  ({\bf b}) a factor of $10^4$, from $0.00001$ to $0.1$ , and  
({\bf c}), a factor of $10^2$ from $0.001$ to $0.1$. In all cases $\eta$ (blue line) dips to a 
minimum then regains its equilibrium value,  $\xi$ (red line) increases to its new equilibrium 
value. The black line shows the bearing angle $\theta$ and dashed grey lines are our 
approximations for $\eta$ and $\xi$ for phases I and II of the transient. 
Vertical scales are logarithmic.
}
\label{fig: 11}
\end{figure}

We begin by developing a solution for phase I of the motion
(where $\xi \ll \xi_0$), by ignoring O($\xi^2$) terms of equations 
(\ref{eq: 6.1}) and (\ref{eq: 6.2}). The latter gives  
\begin{equation}
\label{eq: 6.1.1}
\frac{{\rm d} \eta}{{\rm d} \tilde t} = -\frac{1}{2} \eta^2 \tilde v
\ .
\end{equation}
Taking $\eta = \eta_{\rm in}$, $\tilde t = 0$ at the point of the speed increase, we obtain
\begin{equation}
\label{eq: 6.1.2}
    \eta(\tilde t) = \frac{2 \eta_{\rm in}}{2 + \tilde v \eta_{\rm in} \tilde t}
\ .
\end{equation}
Then from (\ref{eq: 6.1}), we have 
\begin{equation}
\label{eq: 6.1.3}
    \frac{{\rm d} \lambda}{{\rm d} \tilde t} = \frac{1}{2} \tilde v \eta(\tilde t),
\end{equation}
and (\ref{eq: 6.1}) then implies 
\begin{equation}
\label{eq: 6.1.4}
    \xi(\tilde t) = \xi_{\rm in} \exp \left(\frac{\tilde v}{2} \int_0^{\tilde t} \eta({\tilde t}^{\, '})
    {\rm d}\tilde t^{\, '} \right)
\ .
\end{equation}
We now use (\ref{eq: 6.1.2}) in (\ref{eq: 6.1.4}) to estimate $\xi(\tilde t)$:
\begin{equation}
\label{eq: 6.1.5}
    \xi(\tilde t) = \frac{1}{2} \xi_{\rm in} (2 + \tilde v \eta_{\rm in} \tilde t)
\ .
\end{equation}
Equations (\ref{eq: 6.1.2}) (for $\eta(\tilde t)$ decreasing) and (\ref{eq: 6.1.5})
(for $\xi(\tilde t)$ increasing) represent phase I of the transient, 
which (according to (\ref{eq: 6.2})), ends 
when the term $(30-60s)\xi^2(\tilde t)$ becomes comparable to $\tilde v\eta^2/2$. 
We now use these approximations to make an improved estimate of $\eta(\tilde t)$, 
by substituting (\ref{eq: 6.1.5}) into the RHS of (\ref{eq: 6.2}): 
\begin{equation}
\label{eq: 6.1.6}
 \frac{{\rm d} \eta}{{\rm d} \tilde t} \approx 
 -\frac{1}{2} \tilde v \eta(\tilde t)^2 + (30 - 60s) \xi(\tilde t)^2 
 \approx -\frac{2 \tilde v \eta_{\rm in}^2}{(2 + \tilde v \eta_{\rm in} \tilde t)^2} 
 + \frac{1}{4} \xi_{\rm in}^2 (30 - 60s)(2 + \tilde v \eta_{\rm in} \tilde t)^2,
\end{equation}
which we integrate to find an approximation for $\eta$ down to its minimum:
\begin{equation}
\label{eq: 6.1.7}
\eta(\tilde t) \approx 
\frac{(1 - 2 s) \xi_{\rm in}^2 [5 \eta_{\rm in}^3 \tilde v^3 \tilde t^4 + 40 \eta_{\rm in}^2 \tilde v^2 \tilde t^3 
+ 120 \eta_{\rm in} \tilde v \tilde t^2 + 120 \tilde t] +4 \eta_{\rm in}}{2 (2 + \tilde v \eta_{\rm in} \tilde t)}
\ .
\end{equation}
Applying (\ref{eq: 6.1.4}) with its improved approximation to $\eta(\tilde t)$ now gives 
\begin{equation}
\label{eq: 6.1.8}
\xi(\tilde t) \approx \frac{1}{2} \xi_{\rm in} (2 + \eta_{\rm in} \tilde v \tilde t) 
\exp \left[-\frac{5}{16} (1 - 2 s) \eta_{\rm in} \tilde t^2 \xi_{\rm in}^2 \tilde v 
(\eta_{\rm in} \tilde v^2 \tilde t^2 + 8 \tilde v \tilde t + 24) \right]
\ .
\end{equation}
Equation (\ref{eq: 6.1.6}) gives an approximation for the time to the minimum of 
$\eta$ by setting the RHS to zero, that is
\begin{equation}
\label{eq: 6.1.9}
\frac{1}{2} \tilde v \left( \frac{2 \eta_{\rm in}}{2 + \tilde v \eta_{\rm in} \tilde t} \right)^2 
= \frac{1}{4} \xi_{\rm in}^2 (2 + \tilde v \eta_{\rm in} \tilde t)^2 (30 -60s),
\end{equation}
which can be re-arranged as
\begin{equation}
\label{eq: 6.1.10} 
 T^4 = \frac{4 \tilde v \eta_{\rm in}^2}{15 \xi_{\rm in}^2 (1 - 2 s)},
\end{equation}
where $T = 2+{\tilde v}\,{\eta_{\rm in}}\,\underline{\tilde t}$. 
Taking then the real positive root gives an estimate for $\underline{\tilde t}$. 
We compared this estimate  with numerical integration 
of equations (\ref{eq: 2.7}), for three different values of $\xi_{\rm in}$ with $s=0.3904$, 
$\tilde v=0.1$, and  $\eta_0=2.0713$. The results are illustrated in Fig.~\ref{fig: 11}.

When $\xi_{\rm in}=0.000012094$ (which corresponds to steady motion with initial 
speed $v_{\rm in}=0.000000001$), the theoretical minimum of $\eta$ is $0.02260$ 
at $\underline{\tilde t}=1170$, and  the numerical integration gives $\underline{\eta}=0.0203$ 
at $\underline {\tilde t}=1287$ (see figure \ref{fig: 11}(a)).  
When $\xi_{\rm in}=0.0012094$ (corresponding to initial speed $v_{\rm in}=0.00001$), 
the theory predicts a minimum of $\underline{\eta}=0.226$ at $\underline{\tilde t}=108.3$, and 
the numeric values are $\underline{\eta}=0.212$ at $\underline{\tilde t}=116.9$ 
(see figure \ref{fig: 11}(b)). Finally, when $\xi_{\rm in}=0.012094$ 
(corresponding to initial speed $v_{\rm in}=0.001$),  the 
theoretical minimum of $\eta$ is $0.712$ at $\underline{\tilde t}=27.66$, and the numeric 
values are $\underline{\eta}=0.716$ at $\underline{\tilde t}=29.32$ (see figure \ref{fig: 11}(c)).

We can conclude that the theory gives reasonably accurate estimates. 
It is interesting to note that the values of $\underline{\tilde t}$ become large as 
$\xi_{\rm in}$ decreases. Consider what happens if we abruptly accelerate the 
slider from steady motion at speed $v_{\rm in}$ to final speed $v$. If $v/v_{\rm in}$ 
is close to unity, the results of section \ref{sec: 4} showed 
that the transient lasts for a time $t_0\sim L/v_{\rm in}$. If the ratio $v/v_{\rm in}$ 
is large, however, Eq.~(\ref{eq: 6.1.10}) implies that 
\begin{equation}
\label{eq: 6.1.11}
\frac{\underline{t}}{t_0}\sim \left(\frac{v}{v_{\rm in}}\right)^{1/4},
\end{equation}
which is  consistent with the results we have quoted for specific cases.

\subsection{Estimate of energy dissipated}
\label{sec: 6.2}

Now we estimate the excess energy dissipated due to 
an abrupt acceleration. The dissipative force is given by equation (\ref{eq: 2.8}), with 
the coefficients $D_{ij}(\eta)$ specified in Appendix A. The force is greatest in the 
initial phase of the transient, when the gap parameter $\xi$ is still very small. Accordingly,
the force is dominated by the term containing the coefficient $D_{11}$:
\begin{equation}
\label{eq: 6.2.1}
F_x\sim-\frac{\tilde v \ln(1 + \eta(\tilde t))}{\xi(\tilde t) \eta(\tilde t)}
\ .
\end{equation}
Using equations (\ref{eq: 6.1.2}) and (\ref{eq: 6.1.5}), the dissipative force 
during phase I of the transient is 
\begin{equation}
\label{eq: 6.2.2}
F_x(\tilde t)\sim-\frac{\tilde v}{\eta_{\rm in} \xi_{\rm in}} 
\ln \left( 1 + \frac{2 \eta_{\rm in}}{2 + \tilde v \eta_{\rm in} \tilde t} \right)
\ .
\end{equation}
The energy dissipated up to time $\tilde t$ is 
\begin{equation}
\label{eq: 6.2.3}
E = \int_0^{\tilde t} {\rm d}{\tilde t'} \, \tilde v F_x({\tilde t'})
\ .
\end{equation}
Using equation (\ref{eq: 6.2.2}) yields an analytic approximation:
\begin{eqnarray}
\label{eq: 6.2.4}
E&=&-\frac{\tilde v^2}{\eta_{\rm in} \xi_{\rm in}}\int_0^{\tilde  t} {\rm d}{\tilde t'}
\ln \left( 1 + \frac{2 \eta_{\rm in}}{2 + \tilde v \eta_{\rm in} \tilde t'} \right) 
\\
&=&-{\frac {\tilde v}{{{\eta_{\rm in}}}^{2}\xi_{\rm in}} 
\left[ \left( {\eta_{\rm in}}\,{\tilde t}\,{\tilde v}\,+2 {\eta_{\rm in}}+2 \right) 
\ln  \left( {\frac {{\eta_{\rm in}}\,{\tilde t}\,{\tilde v}+2\,{\eta_{\rm in}}+2}
{2 + {\eta_{\rm in}}\,{\tilde t}\,{\tilde v}}} \right) 
-2\,{\eta_{\rm in}}\,\ln  \left(  \frac{2 \eta_{\rm in} + 2}{2 + {\eta_{\rm in}}\,{\tilde t}\,{\tilde v}} \right)
  -2\, \ln  \left( {\eta_{\rm in}}+1 \right) \right] }
 \nonumber
\ .
\end{eqnarray}

\begin{figure*}
\centering
\includegraphics[width=0.4\textwidth]{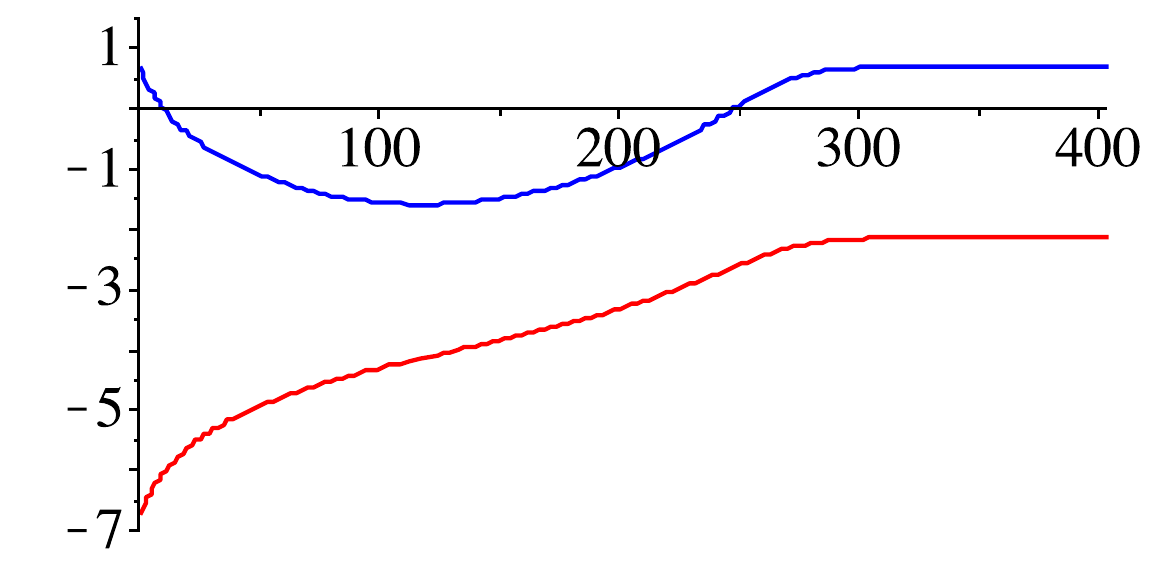}
\includegraphics[width=0.4\textwidth]{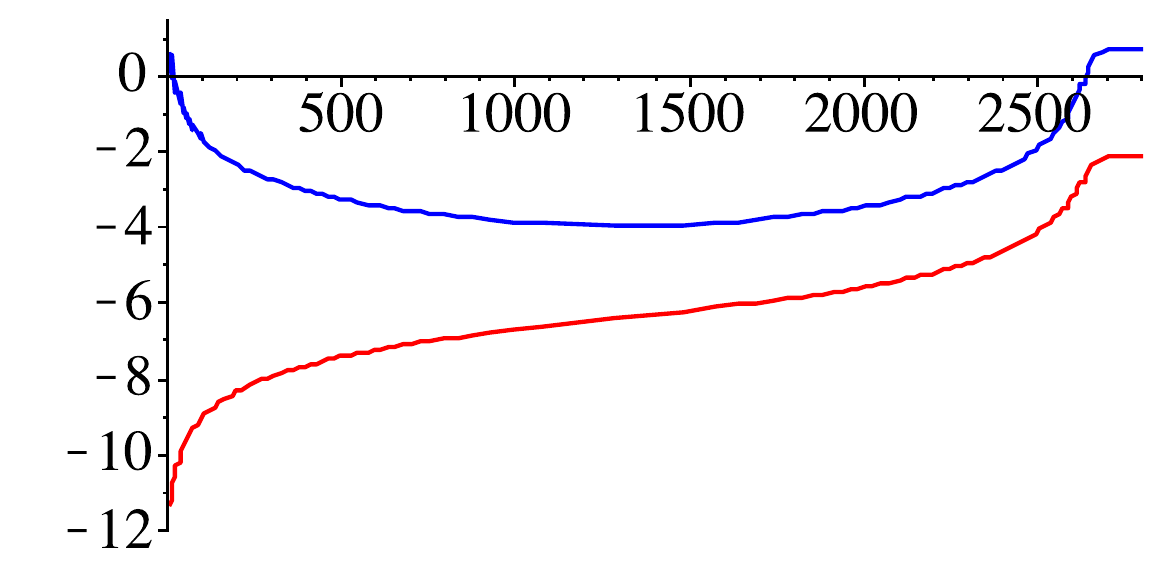}
\includegraphics[width=0.4\textwidth]{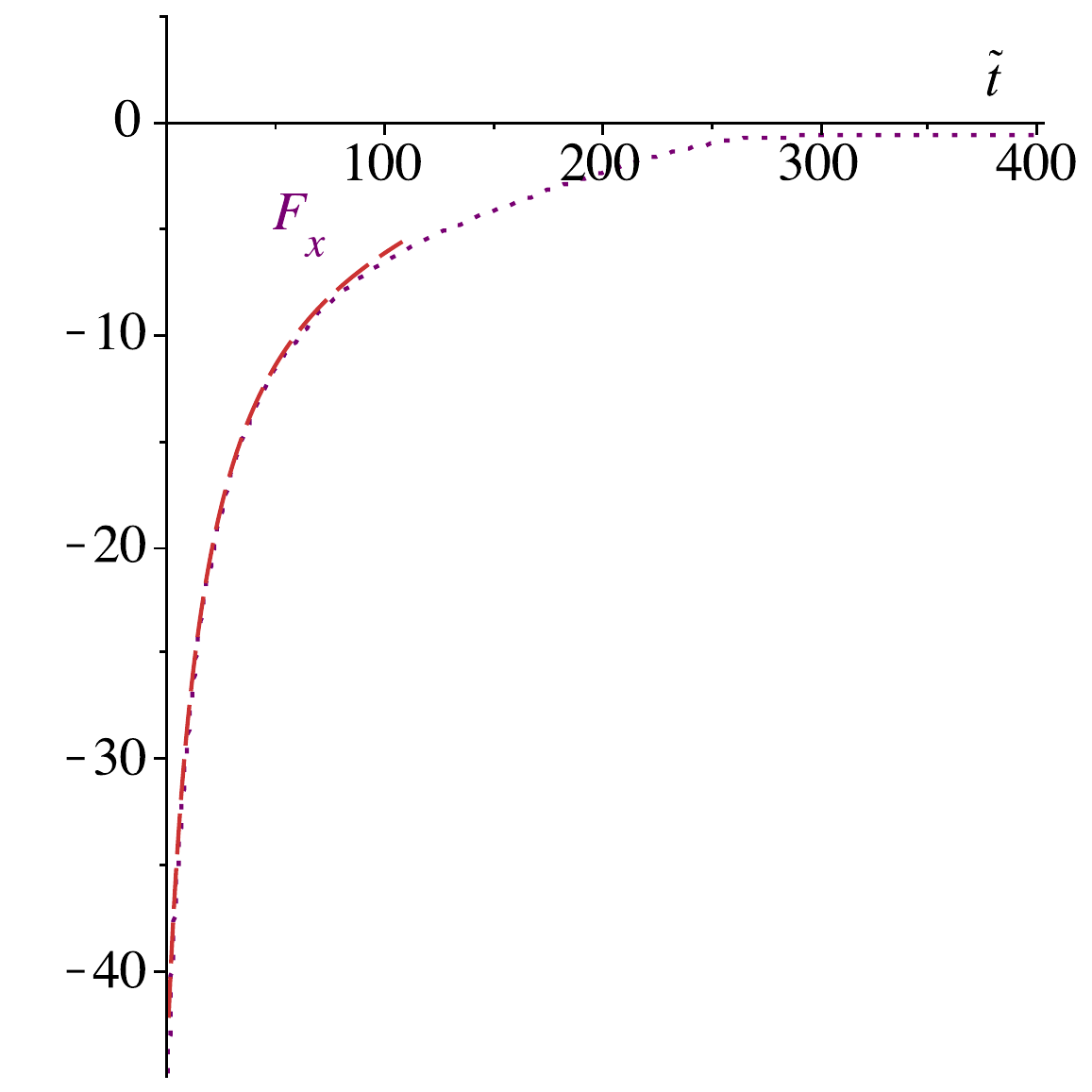}
\includegraphics[width=0.4\textwidth]{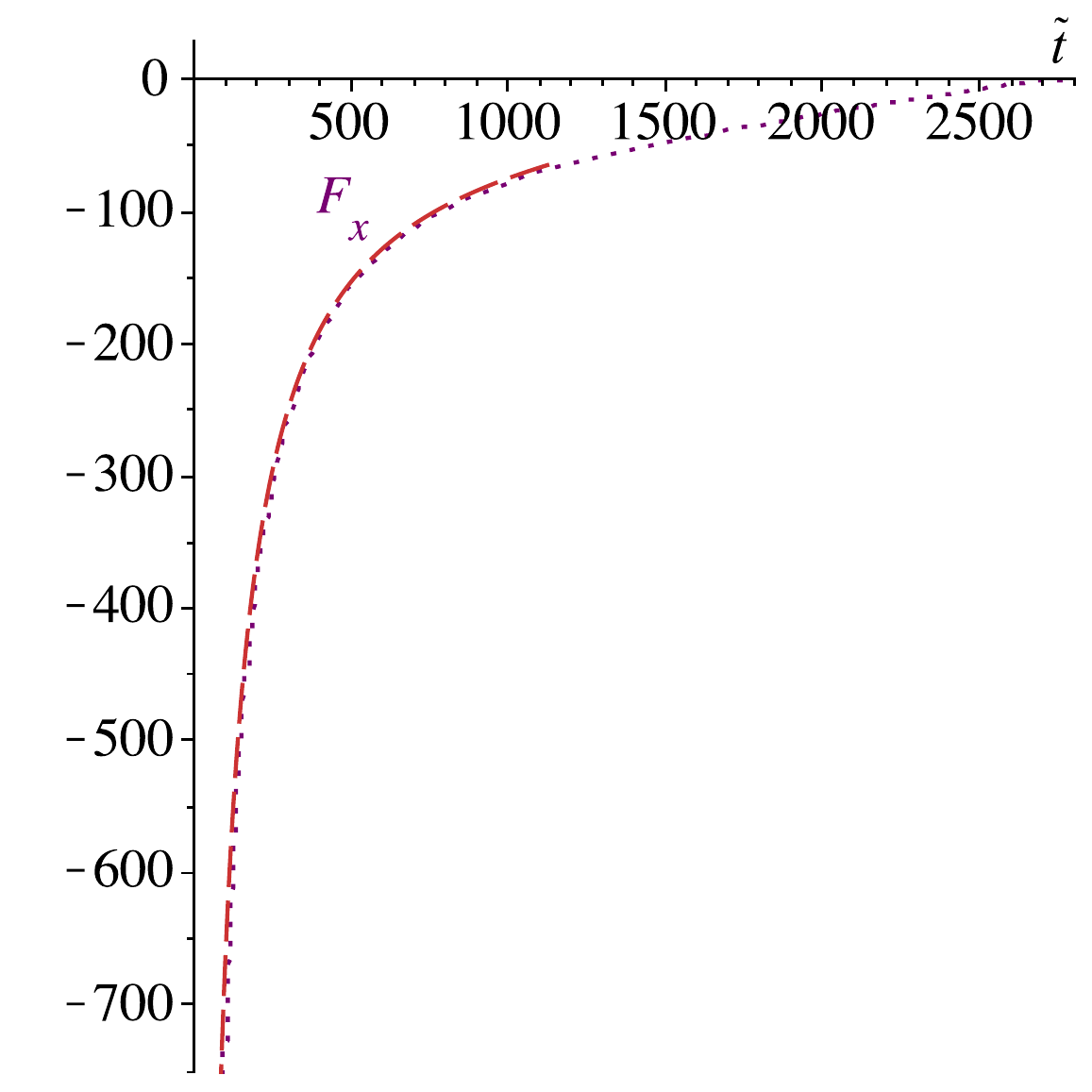}
\caption{Lower panels show that the dissipative force (purple dot is numerical result, 
orange dash is analytic approximation is most significant, well approximated by  (\ref{eq: 6.2.2}), 
during phase I of the motion. Left side plots show a $10^4$ times speed increase, 
right hand plots show a $10^8$ times speed increase. The upper plots show 
$\eta$ (blue) and $\xi$  (red), both log scale, with $\eta$ dipping to a minimum and regaining 
equilibrium, in order to make a comparison with the phases of the motion defined by 
figure \ref{fig: 6.1}. In both cases $s=0.3904$, $\tilde v =0.1$.
}
\label{fig: 14}
\end{figure*}

Figure \ref{fig: 14} illustrates the dissipative force, $F_x$, and $\eta$ during 
the transition from equilibrium at an initial constant speed to equilibrium at a 
much higher constant speed. This decreases very rapidly, indicating 
that the majority of energy dissipated in the transition occurs in the phase 
when $\eta$ is decreasing to its minimum value. 

\begin{figure*}
\centering
\includegraphics[width=0.45\textwidth]{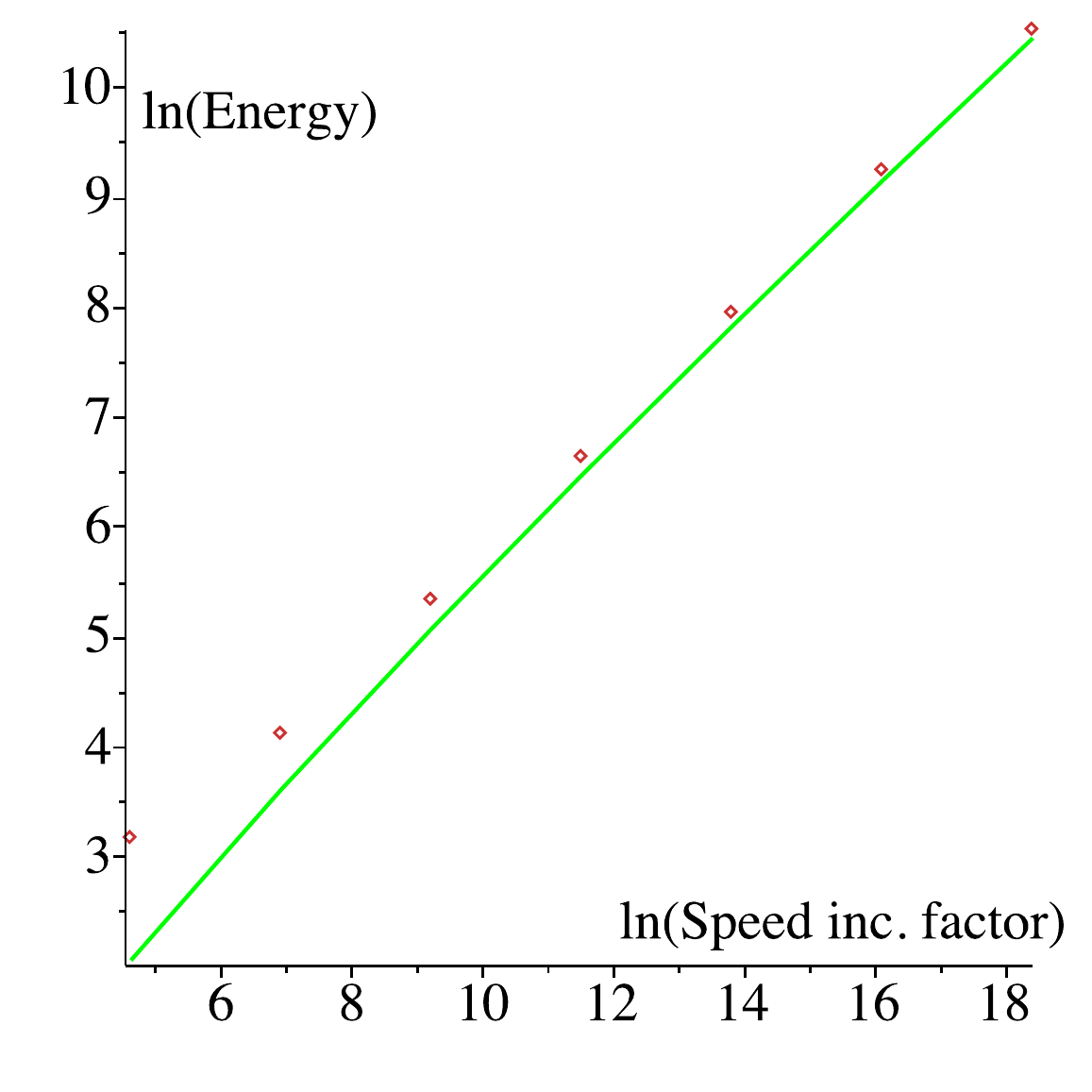}
\includegraphics[width=0.45\textwidth]{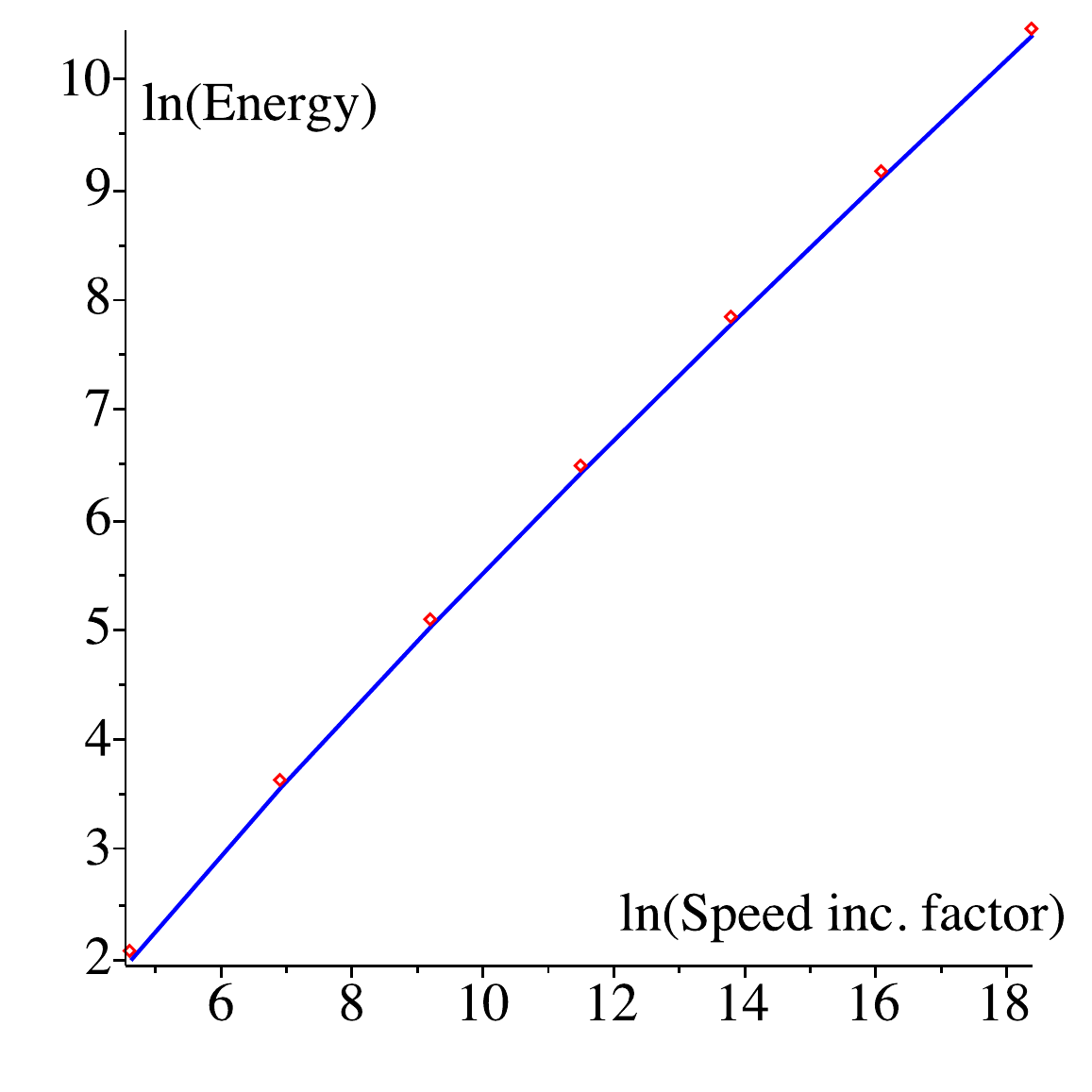}
\caption{
Plots of excess energy dissipated after an abrupt speed increase. 
The left pane shows the total 
energy dissipated at speed increase factors $10^i$, $i=2$ to $8$ (orange diamonds) 
and the energy dissipated to the minimum of $\eta$ (green line), both from the 
numeric model. The right pane shows energy dissipated to the minimum of $\eta$ 
as given by the numeric model at the same speed increase factors (red diamonds) 
and the energy dissipated as predicted by the analytic model (blue line). In both 
cases $s=0.3904$, $\tilde v =0.1$.
}
\label{fig: 15}
\end{figure*}

We made a theoretical estimate of the total energy dissipated 
by evaluating (\ref{eq: 6.2.4}) at the time $\underline{\tilde t}$, estimated from 
equation (\ref{eq: 6.1.10}). Figure \ref{fig: 15} compares this estimate with 
the excess energy dissipation obtained from numerical integration.
It illustrates both the increasing proportion 
of energy dissipated during speed transition as the speed increase factor 
grows (reaching $93 \%$ by a $10^8$ times increase), and the close fit of 
our analytic model of energy dissipation to the minimum of $\eta$ in comparison 
to the numeric model figures.

The foregoing shows that a vast majority of energy dissipated in achieving a very 
large bearing speed increase occurs in the period where $\eta$ is decreasing to 
its minimum. We have a good analytic approximation for energy dissipated in this 
time interval and, using this to approximate the total energy dissipation, we are 
therefore able to look at the different energy dissipations for varying values of pivot 
position, our aim being to comment on whether the constant speed optimum, 
$s=0.3904$ remains a good choice in situations where frequent large speed 
variations will occur. 

We might expect that, because moving the pivot point backwards
increases the steady-state angle parameter $\eta_0$, this will reduce the drag 
experienced upon acceleration by facilitating entrainment of fluid under the bearing.
The effects of changing the pivot point position were numerically significant, {and consistent
with this hypothesis, but not especially dramatic. 
We investigated speed increases by factors of $10^8$, $10^2$, 
and $10$ times up to a final speed $\tilde v = 0.1$. The excess energy dissipation values 
are plotted in Fig.~\ref{fig: 16}, illustrating the reduced energy dissipation 
during the transient when the pivot point is moved closer to the trailing edge. 
The improvement is of the order of $60 \%$, $55 \%$ and $50 \%$ energy saving when 
moving $s$ from $0.3904$ to $0.3$ at speed increase factors of $10^8$, $10^2$, and 
$10$ respectively. However, changing the pivot point causes a large increase of 
the dissipation rate for a bearing moving at constant speed (see Fig.~(\ref{fig: 3.2})), 
so that the speed fluctuations would have to occur very frequently in order for these 
improvements in transient response to become significant.

\begin{figure*}
\centering
\includegraphics[width=0.3\textwidth]{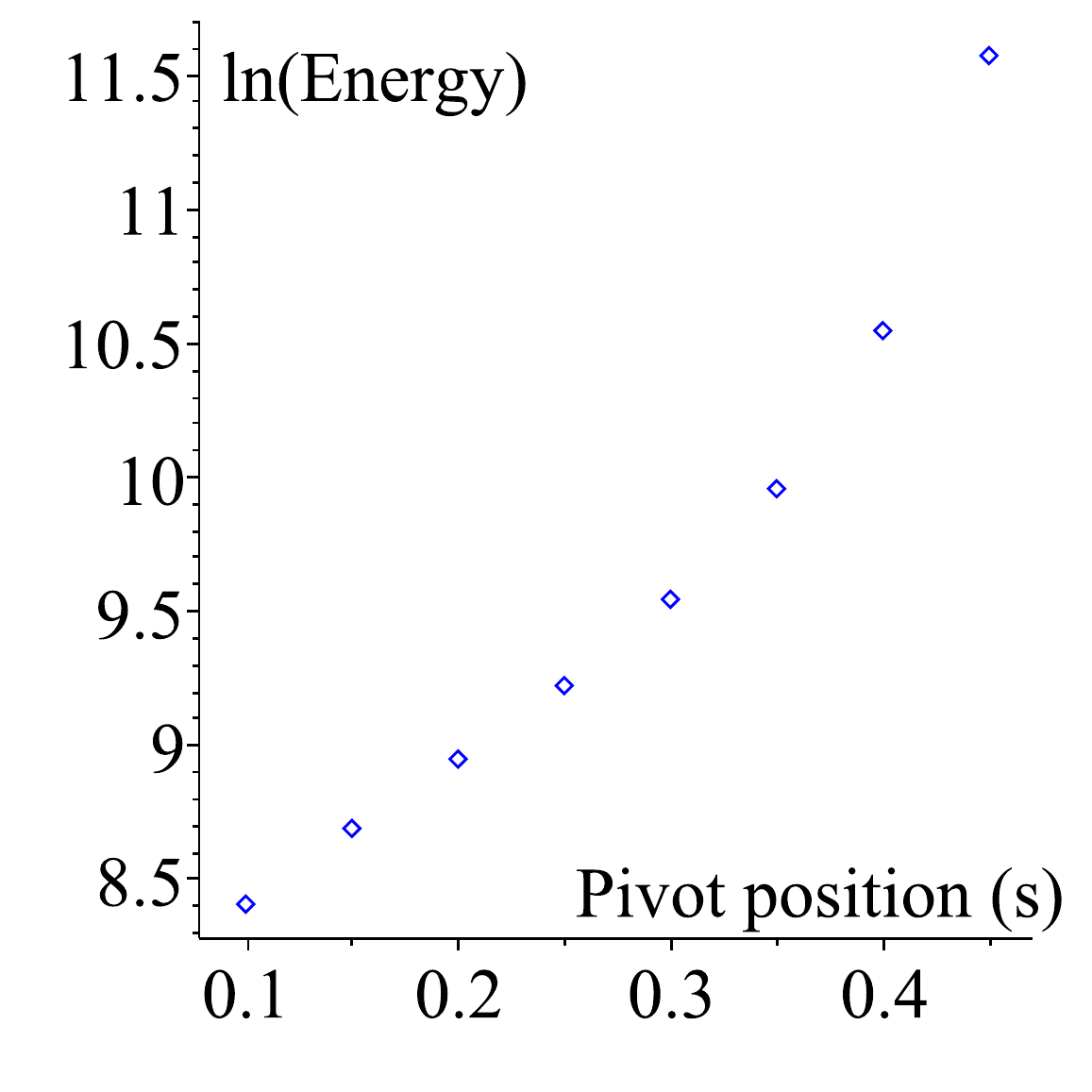}
\includegraphics[width=0.3\textwidth]{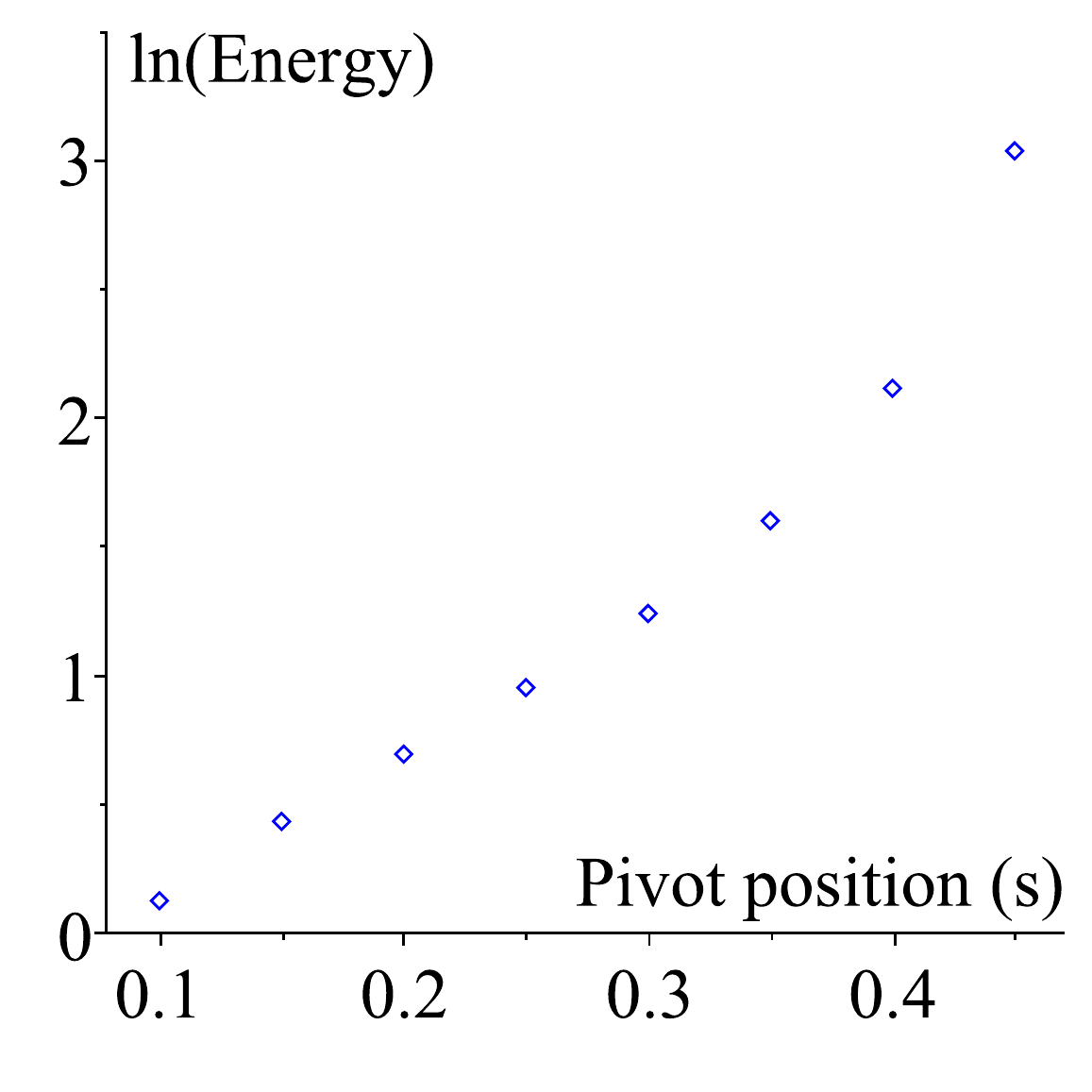}
\includegraphics[width=0.3\textwidth]{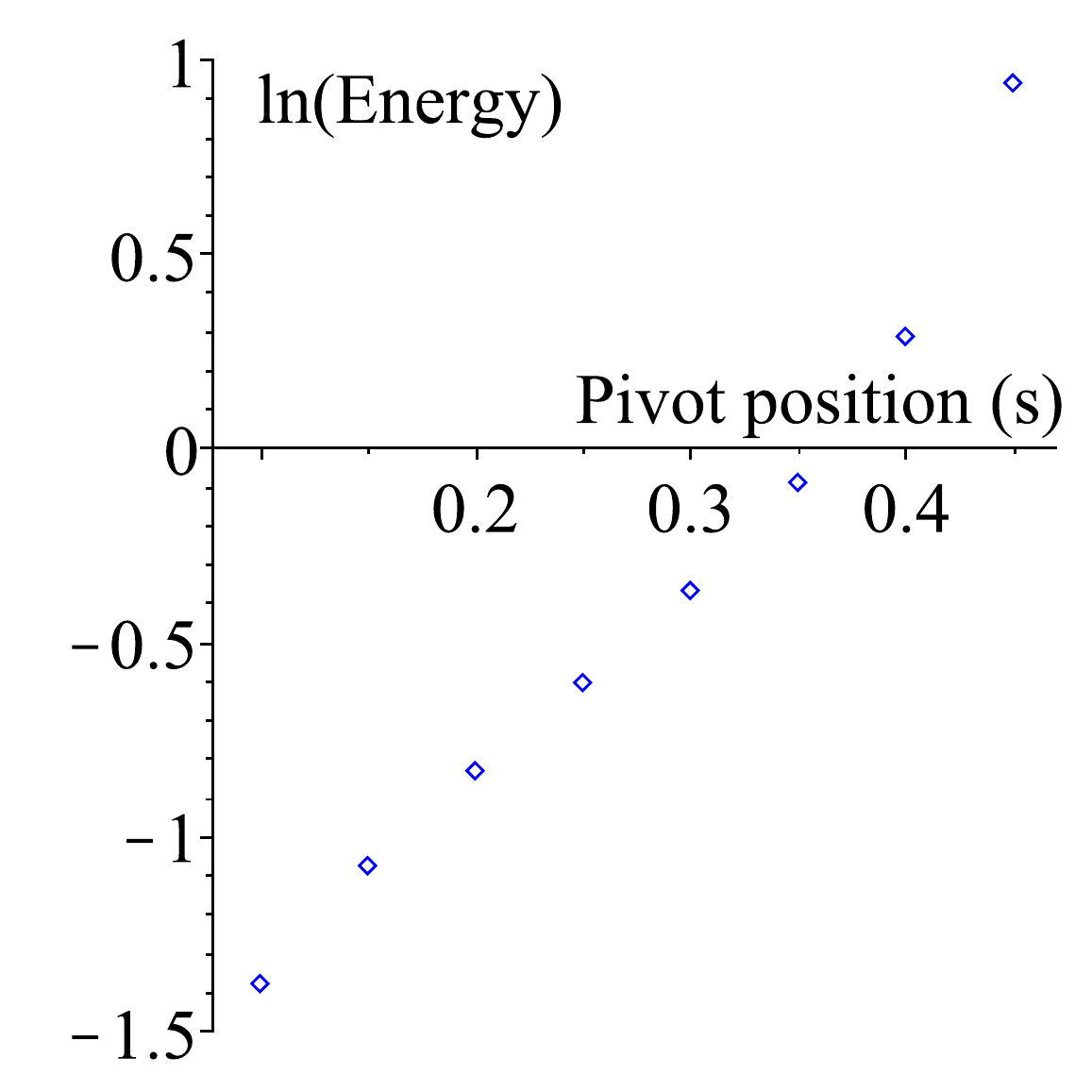}
\caption{
Plot of energy dissipated during equilibrium transition through to the minimum of 
$\eta$, as a function of the pivot position $s$, following an abrupt speed change. 
Left, middle and right panels show speed increases of $10^8$, $10^2$, and $10$ times, 
respectively. $\tilde v =0.1$ is used throughout.
}
\label{fig: 16}
\end{figure*}

\section{Concluding remarks}
\label{sec: 7}

We have obtained the full equations of motion for pivoted slider bearings. 
Our primary motivation was to investigate the transient motion of the slider
after an abrupt increase in its velocity. Our numerical results show that this 
is quite complex, as illustrated in Figs.~\ref{fig: 6.1}-\ref{fig: 11}. 
We were able to analyse this complex transient, and to make an analytical 
approximation for the time-dependence of the dynamical variables 
during its crucial initial stages. We were also able to quantify the additional 
energy dissipation which occurs when the speed of the slider bearing is abruptly
increased.

In order to obtain the equations of motion, we introduced a general notion 
of a \emph{transversion} procedure, to change the dependent variables in 
a dynamical process with linear equations of motion. Using this approach we 
were able to obtain the equations of motion from those of a sinking plate, 
which were obtained in \cite{Wil+23}.

We also determined the sinking motion of the slider after sliding motion ceases, and 
showed that the steady-state solutions investigated in \cite{Mic50} are, in fact, stable.

\section*{Acknowledgments}

We thank Prof. G. W. Milton for pointing out the relevance of the Schur complement to equation 
(\ref{eq: 2.5}).

\appendix

\section*{Appendix A: Matrix coefficients}
\label{AppA}

We list the matrix elements of ${\bf B}$, defined by equation \eqref{eq: 2.3}.
In the following expressions, 
\begin{equation}
\label{eq: A.0}
\psi =\ln(\eta +1) 
\ .
\end{equation}
The elements are:
\begin{eqnarray}
\label{eq: A.1}
B_{11}&=& {\frac { \left( -4\eta-8 \right) \psi +6\eta}{ \left( \eta+2 \right) \eta}}
\nonumber \\
B_{12}&=&{\frac { \left( -6\eta-6 \right) \psi +3\eta^2+6\eta}{ \left( \eta+2 \right) {\eta}^{2}}}
\nonumber \\
B_{13}&=&{\frac { \left( -12\,\eta-18 \right) \psi +3
\,{\eta}^{2}+18\,\eta}{ \left( \eta+2 \right) {\eta}^{3}}}
\nonumber \\
B_{21}&=&{\frac { \left( 6\,\eta+12 \right) \psi -12
\,\eta}{ \left( \eta+2 \right) {\eta}^{2}}} 
\nonumber \\
B_{22}&=&{\frac { \left( 12\,\eta+12 \right) \psi -6
\,{\eta}^{2}-12\,\eta}{ \left( \eta+2 \right) {\eta}^{3}}} 
\nonumber \\
B_{23}&=& {\frac { \left( 24\,\eta+36 \right) \psi -6
\,{\eta}^{2}-36\,\eta}{ \left( \eta+2 \right) {\eta}^{4}}} 
\nonumber \\
B_{31}&=& {\frac { \left( -12\,\eta-18 \right) \psi+3
\,{\eta}^{2}+18\eta}{ \left( \eta+2 \right) {\eta}^{3}}}
\nonumber \\
B_{32}&=& -\frac32\,{\frac { \left(  \left( 2\eta+2 \right) \psi +{\eta}^{2}-2\,\eta \right)  \left(  \left( -2\eta-2
 \right) \psi+{\eta}^{2}+2\eta \right) 
}{ \left( \eta+2 \right) {\eta}^{5}}}
\nonumber \\
B_{33}&=& {\frac {12\, \left( \eta+1 \right) ^{2} \psi^{2}+ \left( -72{\eta}^{2}-96\,\eta
 \right) \psi-3\,{\eta}^{2} \left( {\eta}
^{2}-4\,\eta-28 \right) }{ 2\left( \eta+2 \right) {\eta}^{6}}}
\end{eqnarray}
The matrix elements of ${\bf C}={\bf B}^{-1}$ are:
\begin{eqnarray}
\label{eq: A.2}
C_{11}&=&-\frac {\eta}{\psi}
\nonumber \\
C_{12}&=&{\frac {-{\eta}^{2}}{2\psi}} 
\nonumber \\
C_{13}&=&0 
\nonumber \\
C_{21}&=& {\frac {-{\eta}^{2}}{2\psi}}
\nonumber \\
C_{22}&=&\frac{\left(4\, \left( \eta+1 \right)^{2} \psi^{3}-\left(6\eta^{2}+8\eta \right) \psi^{2}-\eta^{2} \left(\eta^2+8\eta+2 \right) \psi +3\eta^{4}+6\eta^{3} \right) {\eta}^{3}}
{3 \left(  \left( -2\eta-2 \right) \psi +{\eta}^{2}+2\eta \right) \psi  \left( 
 \left( 2\eta+2 \right) \psi^{2}+ \left( {\eta}^{2}+2\eta \right) \psi -4\eta^{2} \right) }
 \nonumber \\
C_{23}&=&\frac {\left(\left( -4\eta-6 \right) \psi+\eta^2+6\eta \right) \eta^5}
 {3((-2\eta-2)\psi+\eta^2+2\eta)  (( 2\eta+2) \psi^2+ (\eta^2+2\eta) \psi-4\eta^2)}
 \nonumber \\
C_{31}&=&{\frac {{\eta}^{3}}{2\psi}} 
\nonumber \\
C_{32}&=&{\frac { \left(  \left( 2\eta+2 \right)  \psi^{2}+ \left( {\eta}^{2}+\eta \right) 
\psi-3\eta^{2} \right) {\eta}^{4}}
{3\psi \left(  \left( 2\eta+2 \right) \psi ^{2}+ \left( {\eta}^{2
}+2\,\eta \right) \psi-4\eta^{2}
 \right) }}
 \nonumber \\
C_{33}&=&
-{\frac {{\eta}^{6}}{ \left( 6\,\eta+6 \right)  \psi^{2}+ \left( 3\eta^{2}+6\eta \right) \psi-12\eta^{2}}}
%
\end{eqnarray}

The matrix elements $D_{ij}(\eta)$ are:
\begin{eqnarray}
\label{eq: A.3}
D_{11}(\eta)&=&-{\frac {\psi}{\eta}}
\nonumber \\
D_{12}(\eta)&=&-\frac {\eta}{2}
\nonumber \\
D_{13}(\eta)&=&0
\nonumber \\
D_{21}(\eta)&=&\frac{\eta}{2}
\nonumber \\
D_{22}(\eta)&=&-\,{\frac { \left( 4\,  \psi  
^{2}{\eta}^{2}-{\eta}^{4}+8\,  \psi
  ^{2}\eta-24\,\psi {\eta}^{2}+4\,{\eta}^
{3}+4\,  \psi   ^{2}-32\,\psi \eta+28\,{\eta}^{2} \right) {\eta}^{3}}{12 \, \left(4\,
  \psi   ^{3}{\eta}^{2}-\psi {\eta}^{4}+8\,  \psi   ^{3}\eta-12\,\psi {\eta}^{3}+4
\,{\eta}^{4}+4\,  \psi   ^{3}-12\,
\psi {\eta}^{2}+8\,{\eta}^{3} \right)}}
\nonumber \\
D_{23}(\eta)&=&{\frac { \left( 4\,\psi \eta-{\eta}^{2}+6\,
\psi -6\,\eta \right) {\eta}^{5}}{3 \, \left( 4\, 
\psi   ^{3}{\eta}^{2}-\psi {\eta}^{4}+8\,  \psi   ^{3
}\eta-12\,\psi {\eta}^{3}+4\,{\eta}^{4}+4\,
  \psi   ^{3}-12\,\psi {\eta}^{2}+8\,{\eta}^{3} \right) }}
\nonumber \\
D_{31}(\eta)&=&-\frac{\eta^2}{2}
\nonumber \\
D_{32}(\eta)&=&{\frac {{\eta}^{4} \left( 2\,\psi \eta+{
\eta}^{2}+2\,\psi -2\,\eta \right) }{12 \, \left( 2\, 
\psi   ^{2}\eta+\psi {\eta}^{2}+2\,  \psi   ^{2
}+2\,\psi \eta-4\,{\eta}^{2} \right)}}
\nonumber \\
D_{33}(\eta)&=&-{\frac {{\eta}^{6}}{3 \, \left(2\,  \psi
  ^{2}\eta+\psi {\eta}^{2}+2\, 
\psi   ^{2}+2\,\psi \eta-4\,{\eta}^{2} \right)}}
\end{eqnarray}

\end{document}